\documentclass[aps,prb,twocolumn,superscriptaddress,longbibliography]{revtex4-2}
\usepackage{amsmath,amssymb}
\usepackage[pdftex]{hyperref,graphicx}
\hypersetup{colorlinks = true, urlcolor = blue, linkcolor = blue, citecolor = blue}
\usepackage{physics}
\usepackage{xcolor}
\usepackage{bm}
\newcommand{\anno}[1]{}

\begin{document}
\title{On-demand large-conductance in trivial zero-bias tunneling peaks in Majorana nanowires}
\author{Haining Pan}
\affiliation{Condensed Matter Theory Center and Joint Quantum Institute, Department of Physics, University of Maryland, College Park, Maryland 20742, USA}
\author{Sankar Das Sarma}
\affiliation{Condensed Matter Theory Center and Joint Quantum Institute, Department of Physics, University of Maryland, College Park, Maryland 20742, USA}

\begin{abstract}
    Motivated by recent experiments that report the almost-generic large-conductance peaks without very extensive fine-tuning, we propose an alternative mechanism through direct theoretical simulations that can explain the large zero-bias conductance peaks being generated on-demand in the nontopological regime in Majorana nanowires by satisfying the following three sufficient conditions: (i) strong potential disorder in the bulk of the nanowire, suppressing the topological regime; (ii) strong suppression of the disorder near the nanowire ends connecting to the tunneling leads, perhaps because of screening by the metallic leads and gates; and (iii) low tunnel barrier strength leading to large tunneling amplitude.  The third condition is typically achieved experimentally by fine-tuning the tunnel barrier and the first condition is generic in all existing nanowires by virtue of considerable sample disorder induced by unintentional random quenched charged impurities. The second condition is likely to apply to many samples since the disorder potential would be typically screened more strongly at the wire ends because of the large metallic tunnel pads used experimentally.  We show that the resultant tunneling conductance manifests large trivial zero-bias peaks almost on demand, and such peaks could be $\sim 2e^2/h$, when appropriately fine-tuned by the tunnel barrier strength and the temperature, as reported experimentally. {Our work not only solves the mystery in recent experiments that the observations of the large zero-bias conductance peaks are generic by proposing a theoretically possible mechanism but also explains why these hypothesized conditions are naturally satisfied in experiments.}
\end{abstract}

\maketitle
\section{Introduction}\label{sec:introduction}
 The search for topological Majorana zero modes in nanowires with superconducting proximity effect is among the most active research areas in physics~\cite{sarma2015majorana,sau2021majorana,lutchyn2018majorana}. The theoretical predictions made in 2010 are precise~\cite{sau2010nonabelian,lutchyn2010majorana,sau2010generic,oreg2010helical}: Take a semiconductor (SM) nanowire, InAs or InSb, with strong spin-orbit (SO) coupling with a nearby superconductor (SC), Al or Nb, inducing proximity effect in the nanowire; then apply a magnetic field along the nanowire to create a Zeeman spin splitting; and for suitable theoretically defined values of spin splitting $V_z$, the chemical potential $\mu$, and the induced proximity SC gap $\Delta$, the nanowire will develop topological SC provided $V_z ^2 > \mu^2 + \Delta ^2$, with $V_{zc}= (\mu^2 + \Delta ^2)^{1/2}$ being the topological quantum phase transition (TQPT) point where the SC gap closes with the system transitioning from a trivial SC to a topological SC. The topological SC with$ V_z > V_{zc}$  comes with a bulk gap that protects emergent non-Abelian Majorana zero modes (MZMs) localized at the wire ends.  If the wire is long enough, so that the two MZMs have little overlap, the system is exponentially topologically protected with effectively isolated anyonic MZMs, leading to the possibility of a limited version of fault-tolerant topological quantum computing using Ising anyons~\cite{nayak2008nonabelian}.

The standard scheme to identify MZMs has been tunneling spectroscopy~\cite{sengupta2001midgap,sau2010nonabelian,flensberg2010tunneling,wimmer2011quantum,law2009majorana}, where MZMs lead to zero-bias conductance peaks (ZBCPs) of the quantized value $2e^2/h$ at $T=0$, although this strict quantization may not apply at nonzero temperatures~\cite{setiawan2017electron}. Early experiments indicated the existence of ZBCPs with small values $\ll 2e^2/h$, which were tentatively claimed to be evidence for MZMs~\cite{das2012zerobias,deng2012anomalous,mourik2012signatures,churchill2013superconductornanowire,finck2013anomalous}. But the very small tunnel conductance values of these ZBCPs (as well as the extremely soft nature of the induced SC gap) made the situation inconclusive. In fact, it was clear that these early nanowire samples all had considerable disorder-induced subgap fermionic states, with the small ZBCPs most likely associated with class D electron antilocalization effects~\cite{liu2012zerobias,bagrets2012class,akhmerov2011quantized,liu2012zerobias,sau2013density}.  More recently, however, experiments have reported large ZBCPs $\sim 2e^2/h$ in hard-induced SC gap situations~\cite{nichele2017scaling,zhang2018quantizeda,zhang2021large,yu2021nonmajorana,pan2020situ,song2021large}, which have created considerable excitement as the possible signatures for MZMs, but the typical ZBCP is not stable (e.g., in magnetic field, gate voltage, or tunnel barrier), and they often arise only with considerable fine-tuning of postselected data.  In addition, the ZBCPs are never seen in simultaneous tunneling from both ends, which is a requirement for the nonlocality of MZMs.  In addition, the necessary Majorana oscillations and the opening of a topological gap are never observed either~\cite{dassarma2012splitting}. It now seems almost certain that these large-conductance ZBCPs are induced by strong disorder in the nanowire and are associated with disorder-induced nontopological subgap Andreev bound states which are experimentally generated by fine-tuning and cherry-picking of large data sets~\cite{pan2020physical,pan2020generic,pan2021disorder,pan2021threeterminal,pan2021crossover,pan2021quantized,dassarma2021disorderinduced,woods2021charge,zeng2021partiallyseparated,ahn2021estimating}.

In the current work we show that in certain situations, such large-conductance trivial ZBCPs may arise almost on-demand in disordered nanowires under certain conditions as has recently been reported in both InAs and InSb nanowires~\cite{song2021large,yu2021nonmajorana}. 
{Our mechanism, which explains the generic zero-bias conductance peaks in an experimentally plausible way, was not addressed in previous studies~\cite{pan2020physical,pan2021threeterminal,pan2021quantized} and was not even possible before because the observations of the generic large zero-bias conductance peaks have only been reported in recent experiments~\cite{yu2021nonmajorana,song2021large}.}
We establish that the requirements for such on-demand large ZBCPs are simply that the strong disorder be suppressed near the wire ends where the tunnel barrier/contact is located along with an ability to adjust the tunnel barrier to rather low values to enhance the conductance to the desired large magnitude.  Since the nanowire disorder is likely to be suppressed near the wire ends by virtue of the metallic tunnel contact-induced screening, we think that we have identified the main physical mechanism leading to almost generic large trivial conductance peaks in the tunneling experiments, {which are also likely to happen in experiments in a natural way without invoking too much fine-tuning.}
Such on-demand large ZBCPs are, however, neither robust nor stable, and their conductance would vary strongly with tunnel barrier, magnetic field, temperature, and chemical potential.  In addition, they would typically show up only for tunneling from one end and not manifest any Majorana oscillations.  We have, therefore, solved the mystery of why some recent experiments find numerous large trivial ZBCPs in nanowire experiments~\cite{song2021large}.

\section{Theory}\label{sec:theory} 
\begin{figure}[ht]
    \centering
    \includegraphics[width=3.4in]{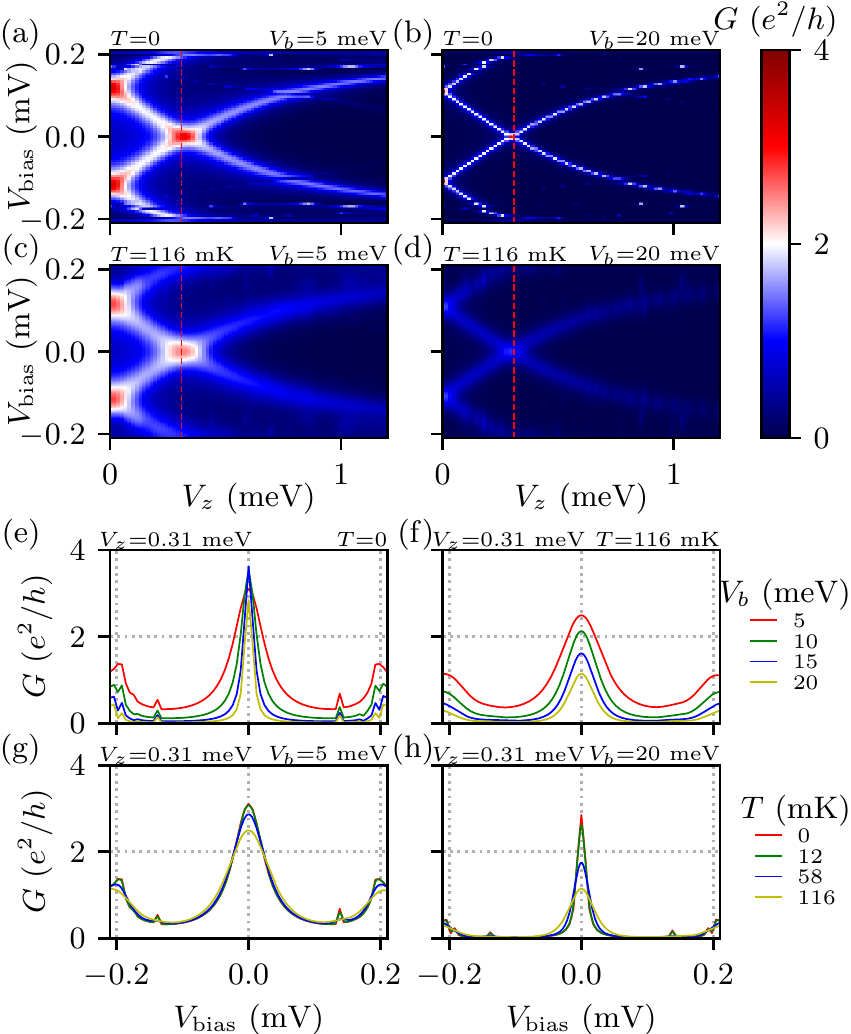}
    \caption{Conductance spectra as a function of the Zeeman field $V_z$ and the bias voltage $V_{\text{bias}}$ in a short wire ($L=1~\mu$m) in the presence of strong disorder in the bulk and strong suppression of the disorder near the ends for (a) temperature $T=0$ and barrier strength $V_b=5$ meV; (b) $T=0$ and $ V_{b} =20$ meV; (c) {$T=116$ mK} and $ V_{b} =5$ meV; (d) {$T=116$ mK} and $ V_{b} = 20 $ meV. Panels (e)-(h) show corresponding line cuts of the trivial ZBCPs at $V_z=0.31$ meV [red dashed lines in panels (a)-(d)] for different temperatures and barrier heights. The standard deviation of Gaussian disorder is $\sigma_\mu=3$ meV in the bulk region and 0.1 meV near the ends ($\sim 0.1~\mu$m). Refer to Sec.~\ref{sec:theory} for other parameters {and Appendix~\ref{App:B} for the wave functions, the local density of states, and the disorder spatial profile}.}
    \label{fig:4}
\end{figure}

We use the minimal one-dimensional one-subband nanowire model solving the Bogoliubov-de Gennes (BdG) equation exactly in the presence of random disorder as described in great detail in Ref.~\cite{pan2020physical}.  This minimal model is known to work very well and reproduce the results of numerically intensive realistic models essentially quantitatively~\cite{pan2021quantized,ahn2021estimating}. The realistic models involve many unknown parameters since the disorder details in the actual complicated hybrid SM-SC platforms with gates and tunneling leads are unknown, and therefore, the minimal model is more appropriate for general theoretical considerations. We refer the reader for more details on the theory {to Appendix~\ref{App:A}} as well as earlier publications~\cite{pan2020physical,pan2021quantized,ahn2021estimating}. The interesting aspect in the current work is that the strength of the random disorder is much higher in the bulk of the wire than near its ends.  We then solve the BdG equation exactly. Using these BdG solutions and the KWANT scattering matrix theory~\cite{groth2014kwant}, we obtain the tunnel conductance as a function of several relevant variables (e.g., temperature, disorder configurations, tunnel barrier, Zeeman splitting, chemical potential, etc.). Unless otherwise stated, we choose the following typical parameters~\cite{lutchyn2018majorana}: the effective mass is 0.015 $m_e$, with $m_e$ being the rest electron mass, the spin-orbit coupling strength is $0.5$ eV\AA{}, the chemical potential in InSb (normal lead) is 1 meV (25 meV), the parent SC gap is 0.2 meV, the SC-SM coupling strength is 0.2 meV, and the wire length is 3 $\mu$m (except for Fig.~\ref{fig:4}, in which it is 1 $\mu$m). Once we obtain the bias-voltage-dependent tunnel conductance at zero temperature $G(V_{\text{bias}};T=0)$, we calculate the tunnel conductance at finite temperature $T$ as $G(V_{\text{bias}};T)=-\int \frac{d f(E-V_{\text{bias}})}{dE} G(V_{\text{bias}},T=0)dE $, where $f(E-V_{\text{bias}};T)$ is the Fermi-Dirac distribution at temperature $T$~\cite{setiawan2017electron}.

\begin{figure}[ht]
    \centering
    \includegraphics[width=3.4in]{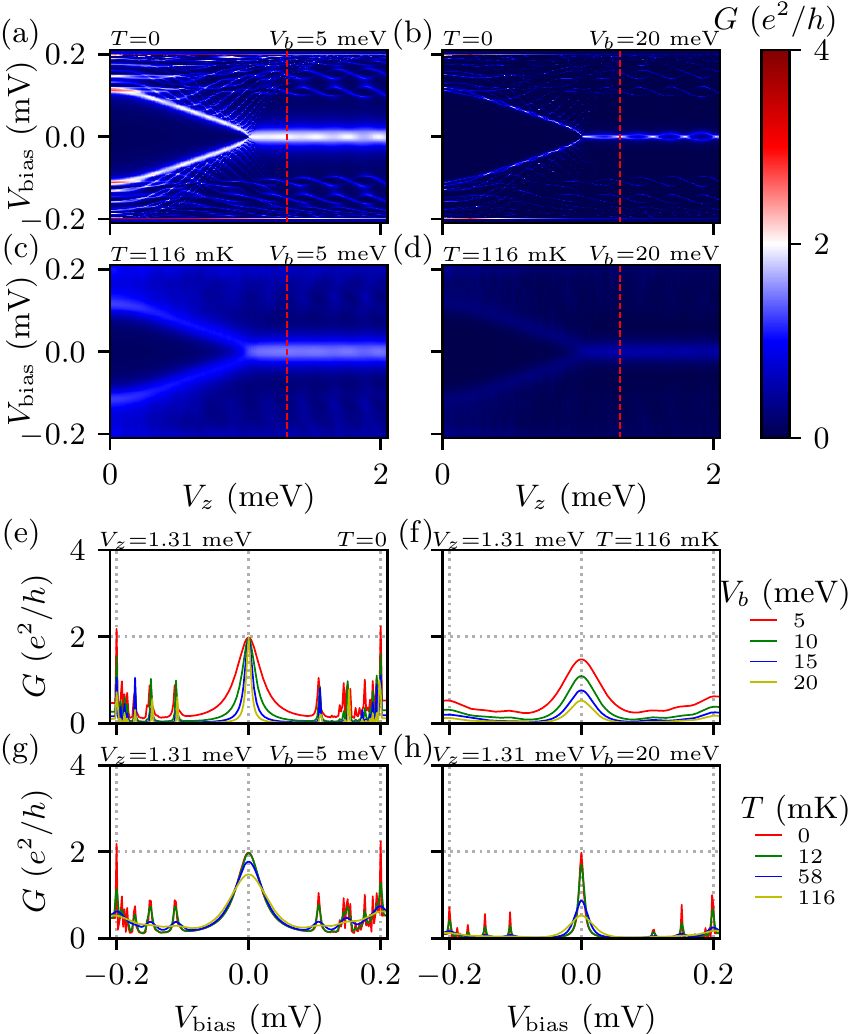}
    \caption{Conductance spectra as a function of the Zeeman field $V_z$ and the bias voltage $V_{\text{bias}}$ in a pristine wire for (a) temperature $T=0$ and barrier strength $V_b=5$ meV; (b) $T=0$ and $ V_{b} =20$ meV; (c) $T=116$ mK and $ V_{b} =5$ meV; (d) $T=116$ mK and $ V_{b} = 20 $ meV. Panels (e)-(h) show corresponding line cuts of the topological ZBCPs at $V_z=1.31$ meV [red dashed lines in panels (a)-(d)] for different temperatures and barrier strengths. Refer to Sec.~\ref{sec:theory} for the parameters {and Appendix~\ref{App:B} for the wave functions and the local density of states}.}
    \label{fig:1}
\end{figure}

\begin{figure*}[ht]
    \centering
    \includegraphics[width=6.8in]{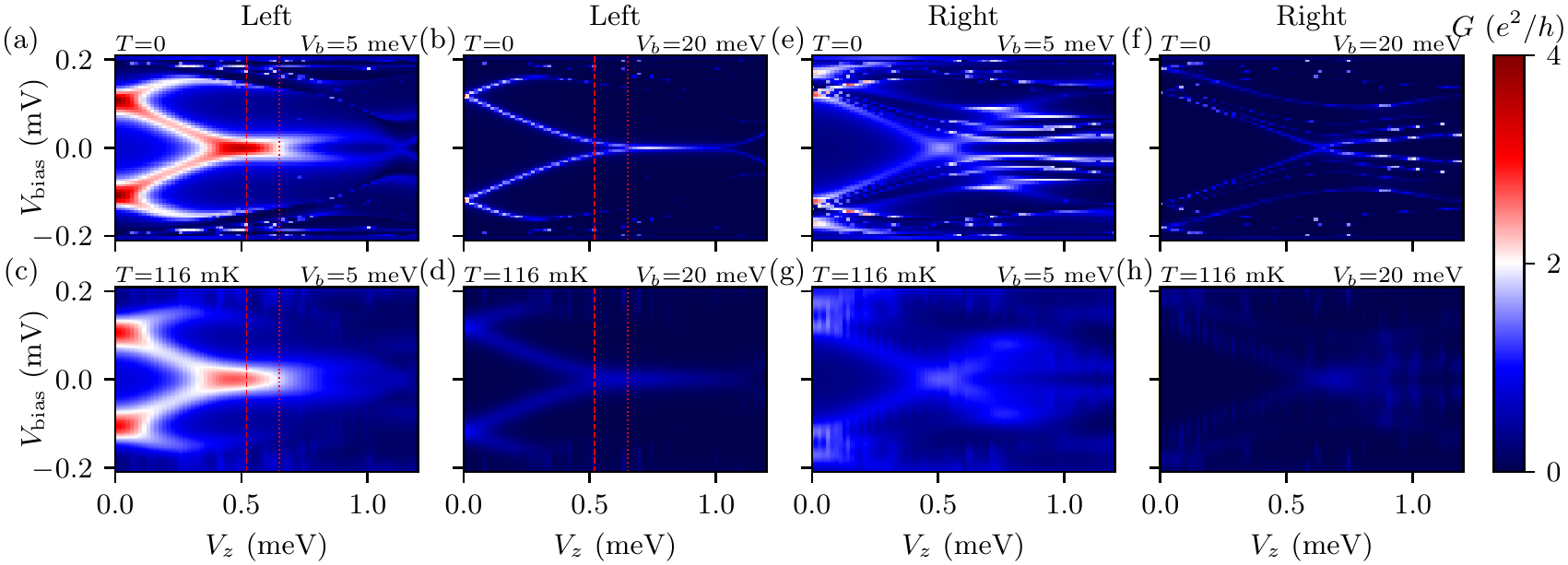}
    \caption{Conductance spectra as a function of the Zeeman field $V_z$ and the bias voltage $V_{\text{bias}}$ in the presence of strong disorder in the bulk and strong suppression of the disorder near the ends for (a) temperature $T=0$ and barrier strength $V_b=5$ meV; (b) $T=0$ and $ V_{b} =20$ meV; (c) {$T=116$ mK} and $ V_{b} =5$ meV; (d) {$T=116$ mK} and $ V_{b} = 20 $ meV. The ZBCPs here are in the trivial regime. Panels (e)-(h) are corresponding conductance spectra measured from the other (right) end. The standard deviation of Gaussian disorder is $\sigma_\mu=3$ meV in the bulk region and 0.5 meV near the ends ($\sim 0.1~\mu$m).  Refer to Sec.~\ref{sec:theory} for other parameters {and Appendix~\ref{App:B} for the wave functions, the local density of states, and the disorder spatial profile}.}
    \label{fig:2}
\end{figure*}

\begin{figure}[ht]
    \centering
    \includegraphics[width=3.4in]{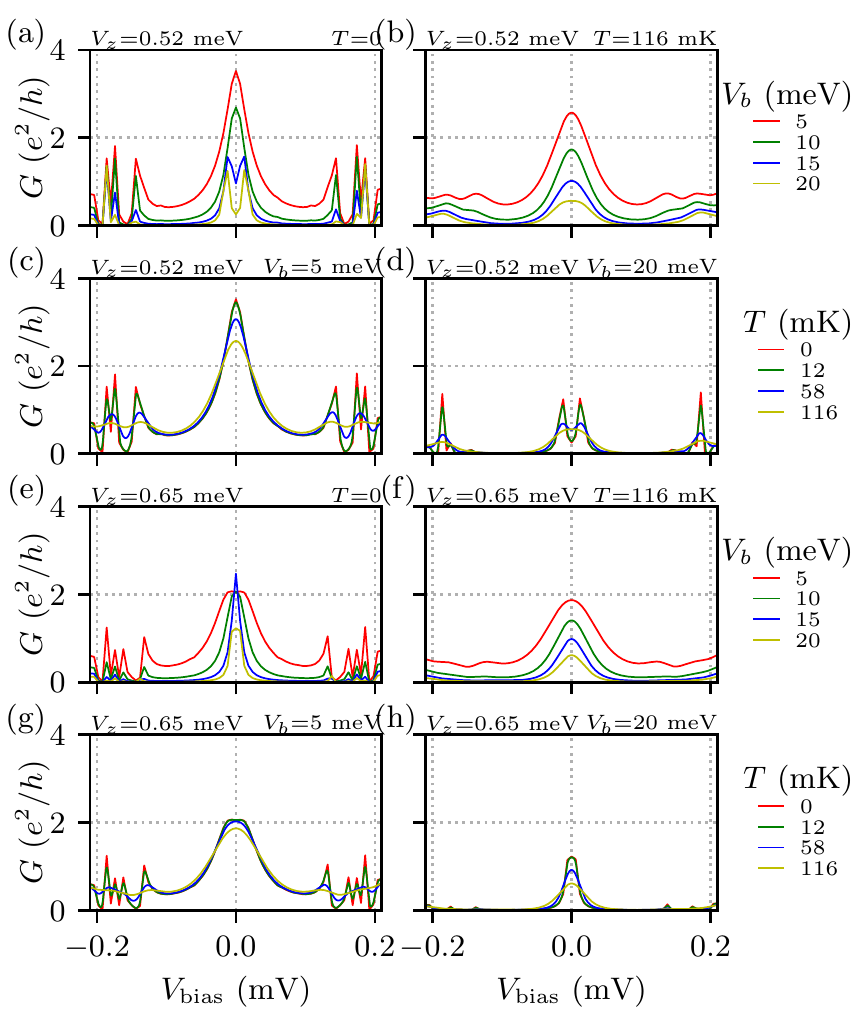}
    \caption{Line cuts of the trivial ZBCPs at a fixed Zeeman field for different temperatures and barrier strengths. Panels (a)-(d) show the line cuts at $V_z=0.52$ meV [red dashed lines in Figs.~\ref{fig:2}(a)-\ref{fig:2}(d)]. Panels (e)-(h) show the line cuts at $V_z=0.65$ meV [red dotted lines in Figs.~\ref{fig:2}(a)-\ref{fig:2}(d)].}
    \label{fig:3}
\end{figure}

\section{Results}\label{sec:results}
 In order to benchmark our results, we start with the pristine wire to study the generic effects of the tunnel barrier and the temperature on the conductance as shown in Fig.~\ref{fig:1}. In Fig.~\ref{fig:1}(a), we present the conductance spectrum of a pristine wire with topological ZBCPs showing beyond the TQPT point ($V_{zc}=1.02$ meV) at zero temperature and low barrier strength ($V_b=5$ meV). As we increase the barrier strength to $V_b=20$ meV in Fig.~\ref{fig:1}(b) while keeping zero temperature, we find that the topological ZBCP becomes sharper, and Majorana oscillations are more salient consequently. However, the conductances of subgap intrinsic Andreev bound states (ABS)~\cite{huang2018metamorphosis} below the TQPT, showing the gap-closing feature, become smaller in Fig.~\ref{fig:1}(b). Therefore, it indicates that the barrier strength does not affect the peak value of topological ZBCPs at zero temperature: it only lowers the conductance peaks of nontopological states. This is also directly manifested in Fig.~\ref{fig:1}(e), where we plot the line cuts of topological ZBCPs at a fixed Zeeman field [$V_z=1.31$ meV indicated by the red dashed line in Fig.~\ref{fig:1}(a)] for different tunnel barrier strengths varying from 5 to 20 meV, and we find that the peaks always stick to the quantized conductance of $2e^2/h$ at zero temperature. 

The finite temperature always suppresses the topological ZBCPs below the quantized conductance of $2e^2/h$, particularly when $T$ is larger than the tunneling energy. For instance, in Fig.~\ref{fig:1}(f), where we increase the temperature to {$T=116$ mK}, topological ZBCPs are no longer quantized, and they decrease as we increase the tunnel barrier strength. We show a comparison between $T=0$  in Fig.~\ref{fig:1}(a) and finite $T$  in Fig.~\ref{fig:1}(c) to emphasize as a benchmark that the topological conductance spectrum is weakened and broadened everywhere by finite temperatures. Thus, all peaks manifest larger linewidths and lower peak values; therefore, the details of subgap states become indiscernible.
In Fig.~\ref{fig:1}(d), we study the joint effect of the finite barrier strength and the finite temperature, and we find that they suppress all tunneling signals in the conductance spectrum, leaving very faint topological ZBCPs and a nearly zero conductance background. We also show the line cuts of conductances in Fig.~\ref{fig:1}(g) to visualize the thermal broadening effect as the temperature increases. From $T=0$ (red line) to {$T=12$ mK} (green line), the two lines overlap, which means the conductance peak does not change too much near zero temperature (i.e., $T$ below tunneling energy). However, as the temperature increases above the tunneling energy to {$T=116$ mK} (yellow line), the conductance peak of the topological ZBCP drops to $\sim 1.5e^2/h$. For the large barrier strength, we present Fig.~\ref{fig:1}(h) and find that the conductance peak drops even more quickly compared to the low barrier strength in Fig.~\ref{fig:1}(g), because the tunneling energy is now exponentially lower than $T$. Note that a finite temperature of {$T=116$ mK (10 $\mu$eV)}, which is less than 10\% of the induced SC gap ($\sim 0.12$ meV), can make the conductance peak drop by a large factor of 75\% [Fig.~\ref{fig:1}(h)] from the original quantized value of $2e^2/h$.  This implies that it is unlikely to observe the quantized value of $2e^2/h$ in experiments even if one manages to measure the conductance of the real topological ZBCP due to the thermal broadening. 

After establishing the benchmark of the finite barrier strength and the temperature in the pristine wire, we introduce our protocol that generates the on-demand large trivial ZBCP in disordered nanowires. The prerequisites are a strong bulk potential disorder and a suppression of the disorder near the nanowire ends. This particular configuration of disorder effectively creates quantum dots near the ends and thus can make the fermionic ABSs localized near the ends, which will then appear as large trivial ZBCPs. We present a representative example in Fig.~\ref{fig:2}(a), which shows a trivial ZBCP with a large conductance of $3e^2/h$. Because the trivial ZBCP is not invariant under the variations in barrier strength, we can manipulate the conductance peak on demand by adjusting the barrier strength as well as the temperature. In Figs.~\ref{fig:2}(b)-\ref{fig:2}(d), we change the conductance of trivial ZBCPs from $2.5e^2/h$ [Fig.~\ref{fig:2}(c)] to $ 0.5e^2/h$ [Fig.~\ref{fig:2}(d)] as we tune the barrier strength and the temperature within the same disorder realization. 
In Figs.~\ref{fig:2}(e)-\ref{fig:2}(h), we present conductances measured from the other end (right) corresponding to Figs.~\ref{fig:2}(a)-\ref{fig:2}(d), {where there are no signatures of zero-bias peaks (see the wave functions and the local density of states in Fig.~\ref{fig:App1})}. Due to the trivial origin of these ZBCPs, the conductances from both ends lack the nonlocal correlation, resembling the experimentally observed ZBCPs.

In addition to the false-color plots of conductances, we also provide the line cuts at a fixed Zeeman field [$V_z=0.52$ meV indicated by red dashed lines in Fig.~\ref{fig:2}] in Figs.~\ref{fig:3}(a)-\ref{fig:3}(d). In Fig.~\ref{fig:3}(a), we notice that the shape of the ZBCP can be tuned into a zero-bias conductance dip (ZBCD) when we increase the barrier strength. Here, the ZBCD is essentially two symmetric side peaks close to each other, which arise from a pair of low-lying trivial ABSs. Therefore, they only appear if the barrier strength is sufficiently high because a high barrier strength suppresses the background conductance between the two side peaks. We show in Fig.~\ref{fig:3}(c) that a small barrier strength fails to manifest a transformation from the ZBCP to the ZBCD. Thus, the transmutation between the ZBCP and the ZBCD, as well as their conductance magnitudes, is easily experimentally tuned by the tunnel barrier~\cite{song2021large}. 

Besides the high tunnel barrier strength, the existence of the ZBCD also needs a low temperature. In Fig.~\ref{fig:3}(b), we increase the temperature from zero to {$T=116$ mK}, finding that all ZBCDs in Fig.~\ref{fig:3}(a) now become ZBCPs. This is because of larger thermal broadening, which could combine two side peaks into one ZBCP, as manifested in Fig.~\ref{fig:3}(d). In Fig.~\ref{fig:3}(d), the ZBCD gradually disappears as the two side peaks merge into one zero-bias peak when the temperature increases, resembling the experimentally observed ZBCDs in Ref.~\onlinecite{song2021large}. Thus, ZBCDs disappear with increasing temperature and/or decreasing tunnel barrier, showing that they represent unstable trivial features.

In addition, the ZBCD, being a fine-tuned feature, is unstable to the applied field. In Fig.~\ref{fig:3}(e), we slightly change the Zeeman field to $V_z=0.65$ meV (red dotted lines in Fig.~\ref{fig:2}), and find that all ZBCDs disappear in comparison with Fig.~\ref{fig:3}(a). At this Zeeman field, the ZBCD cannot exist regardless of the barrier strength and the temperature as shown in Figs.~\ref{fig:3}(e)-\ref{fig:3}(h). 

Although all conductances above from Fig.~\ref{fig:1} to Fig.~\ref{fig:3} are obtained in a wire of 3 $\mu$m, our protocol also works for the short wire, which is closer to the experimental situation. To show this, we consider a shorter wire of 1 $\mu$m in Fig.~\ref{fig:4} with strong potential disorder in the bulk region and suppression of the potential disorder near the nanowire ends. 
{The conductance spectra shown in Fig.~\ref{fig:4} are measured at the right end in the presence of a specific disorder profile [shown in Fig.~\ref{fig:App1}(i)] because the left end does not manifest large conductance of zero-bias peaks.}
In Fig.~\ref{fig:4}(a), we generate the trivial ZBCP with on-demand large conductance with a low tunnel barrier. By increasing the barrier strength and the temperature, we find that the trivial ZBCPs vary drastically as shown in Figs.~\ref{fig:4}(b)-\ref{fig:4}(d). From the line cuts of trivial ZBCPs in Figs.~\ref{fig:4}(e)-\ref{fig:4}(h), we see that the zero-bias conductance declines from $3.5e^2/h$ [blue line in Fig.~\ref{fig:4}(e)] to $e^2/h$ [yellow line in Fig.~\ref{fig:4}(h)] when the barrier strength increases from 5 to 20 meV and the temperature increases from zero to {$T=116$ mK}. All these results manifest the efficacy of our protocol in creating the on-demand trivial ZBCP with large conductance.

\section{Conclusion}\label{sec:conclusion}
We show in this work that there are experimentally relevant physical situations where large zero-bias peaks in the tunneling spectra may be somewhat abundantly present in spite of being entirely trivial in origin.  Such peaks can be made to coincide with the quantized conductance $2e^2/h$  simply by adjusting the tunnel barrier strength (or alternatively perhaps the temperature at a fixed tunnel barrier).  One may even see in some situations transitions in the trivial regions between large-conductance peaks and large-conductance dips in the tunnel spectra~\cite{song2021large}.  Our results show that such trivial tunnel spectra arise from strong disorder in the bulk modulated by the weakening of the disorder at the wire ends (making the ends behave like they are quantum dots), {which is not only theoretically possible but also experimentally plausible because the large disorder induced by the bulk region is always generic and inevitable and weak disorder is suppressed by the screening of the metallic gate.} 

{Although our theory is not guaranteed to be the unique mechanism that can happen in realistic experiments because we simply cannot rule out other possibilities, we show that such a simple yet experimentally plausible hypothesis can lead to the theoretically simulated generic ZBCPs on demand. Therefore, the nutshell of the current work is that on-demand trivial ZBCPs may arise without fine-tuning under certain experiment conditions, and we propose a very likely mechanism that the screening-induced appearance of artificial quantum dots at the ends of the wire plays a substantial role in creating large-conductance peaks on demand which do not manifest much robustness.}

Therefore, experimentalists should be skeptical of any zero-bias peaks not robust to the tunnel barrier, the magnetic field, and the chemical potential (i.e., gate voltage).  In addition, the manifestation of a gap opening and the occurrence of peaks in tunneling from both ends are the minimal requirements for the existence of Majorana zero modes.  In fact, any topological Majorana mode is unlikely to manifest $2e^2/h$ quantization because in most situations finite temperature should suppress the peak value--- what is important is not the peak conductance, but the robust stability of the conductance feature to experimental tuning parameters and the occurrence of the feature for tunneling from both ends.  Obtaining ``quantization'' by fine-tuning the tunnel barrier is perhaps better avoided in the search for topological Majorana zero modes. What is necessary are simultaneous ZBCPs from both ends which are stable to parameter variations independent of the actual conductance values.

This work is supported by the Laboratory for Physical Sciences. We also acknowledge the University of Maryland High-Performance Computing Cluster (HPCC).

\bibliography{Paper_On-demand.bib}

\begin{thebibliography}{42}%
\makeatletter
\providecommand \@ifxundefined [1]{%
 \@ifx{#1\undefined}
}%
\providecommand \@ifnum [1]{%
 \ifnum #1\expandafter \@firstoftwo
 \else \expandafter \@secondoftwo
 \fi
}%
\providecommand \@ifx [1]{%
 \ifx #1\expandafter \@firstoftwo
 \else \expandafter \@secondoftwo
 \fi
}%
\providecommand \natexlab [1]{#1}%
\providecommand \enquote  [1]{``#1''}%
\providecommand \bibnamefont  [1]{#1}%
\providecommand \bibfnamefont [1]{#1}%
\providecommand \citenamefont [1]{#1}%
\providecommand \href@noop [0]{\@secondoftwo}%
\providecommand \href [0]{\begingroup \@sanitize@url \@href}%
\providecommand \@href[1]{\@@startlink{#1}\@@href}%
\providecommand \@@href[1]{\endgroup#1\@@endlink}%
\providecommand \@sanitize@url [0]{\catcode `\\12\catcode `\$12\catcode
  `\&12\catcode `\#12\catcode `\^12\catcode `\_12\catcode `\%12\relax}%
\providecommand \@@startlink[1]{}%
\providecommand \@@endlink[0]{}%
\providecommand \url  [0]{\begingroup\@sanitize@url \@url }%
\providecommand \@url [1]{\endgroup\@href {#1}{\urlprefix }}%
\providecommand \urlprefix  [0]{URL }%
\providecommand \Eprint [0]{\href }%
\providecommand \doibase [0]{https://doi.org/}%
\providecommand \selectlanguage [0]{\@gobble}%
\providecommand \bibinfo  [0]{\@secondoftwo}%
\providecommand \bibfield  [0]{\@secondoftwo}%
\providecommand \translation [1]{[#1]}%
\providecommand \BibitemOpen [0]{}%
\providecommand \bibitemStop [0]{}%
\providecommand \bibitemNoStop [0]{.\EOS\space}%
\providecommand \EOS [0]{\spacefactor3000\relax}%
\providecommand \BibitemShut  [1]{\csname bibitem#1\endcsname}%
\let\auto@bib@innerbib\@empty
\bibitem [{\citenamefont {Sarma}\ \emph {et~al.}(2015)\citenamefont {Sarma},
  \citenamefont {Freedman},\ and\ \citenamefont {Nayak}}]{sarma2015majorana}%
  \BibitemOpen
  \bibfield  {author} {\bibinfo {author} {\bibfnamefont {S.~D.}\ \bibnamefont
  {Sarma}}, \bibinfo {author} {\bibfnamefont {M.}~\bibnamefont {Freedman}},\
  and\ \bibinfo {author} {\bibfnamefont {C.}~\bibnamefont {Nayak}},\ }\bibfield
   {title} {\bibinfo {title} {Majorana zero modes and topological quantum
  computation},\ }\href {https://doi.org/10.1038/npjqi.2015.1} {\bibfield
  {journal} {\bibinfo  {journal} {npj Quantum Information}\ }\textbf {\bibinfo
  {volume} {1}},\ \bibinfo {pages} {15001} (\bibinfo {year}
  {2015})}\BibitemShut {NoStop}%
\bibitem [{\citenamefont {Sau}\ and\ \citenamefont
  {Tewari}(2021)}]{sau2021majorana}%
  \BibitemOpen
  \bibfield  {author} {\bibinfo {author} {\bibfnamefont {J.}~\bibnamefont
  {Sau}}\ and\ \bibinfo {author} {\bibfnamefont {S.}~\bibnamefont {Tewari}},\
  }\bibfield  {title} {\bibinfo {title} {From {{Majorana}} fermions to
  topological quantum computation in semiconductor/superconductor
  heterostructures},\ }\href {http://arxiv.org/abs/2105.03769} {\bibfield
  {journal} {\bibinfo  {journal} {arXiv:2105.03769}\ } (\bibinfo {year}
  {2021})}\BibitemShut {NoStop}%
\bibitem [{\citenamefont {Lutchyn}\ \emph {et~al.}(2018)\citenamefont
  {Lutchyn}, \citenamefont {Bakkers}, \citenamefont {Kouwenhoven},
  \citenamefont {Krogstrup}, \citenamefont {Marcus},\ and\ \citenamefont
  {Oreg}}]{lutchyn2018majorana}%
  \BibitemOpen
  \bibfield  {author} {\bibinfo {author} {\bibfnamefont {R.~M.}\ \bibnamefont
  {Lutchyn}}, \bibinfo {author} {\bibfnamefont {E.~P. A.~M.}\ \bibnamefont
  {Bakkers}}, \bibinfo {author} {\bibfnamefont {L.~P.}\ \bibnamefont
  {Kouwenhoven}}, \bibinfo {author} {\bibfnamefont {P.}~\bibnamefont
  {Krogstrup}}, \bibinfo {author} {\bibfnamefont {C.~M.}\ \bibnamefont
  {Marcus}},\ and\ \bibinfo {author} {\bibfnamefont {Y.}~\bibnamefont {Oreg}},\
  }\bibfield  {title} {\bibinfo {title} {Majorana zero modes in
  superconductor\textendash semiconductor heterostructures},\ }\href
  {https://doi.org/10.1038/s41578-018-0003-1} {\bibfield  {journal} {\bibinfo
  {journal} {Nature Reviews Materials}\ }\textbf {\bibinfo {volume} {3}},\
  \bibinfo {pages} {52} (\bibinfo {year} {2018})}\BibitemShut {NoStop}%
\bibitem [{\citenamefont {Sau}\ \emph {et~al.}(2010{\natexlab{a}})\citenamefont
  {Sau}, \citenamefont {Tewari}, \citenamefont {Lutchyn}, \citenamefont
  {Stanescu},\ and\ \citenamefont {Das~Sarma}}]{sau2010nonabelian}%
  \BibitemOpen
  \bibfield  {author} {\bibinfo {author} {\bibfnamefont {J.~D.}\ \bibnamefont
  {Sau}}, \bibinfo {author} {\bibfnamefont {S.}~\bibnamefont {Tewari}},
  \bibinfo {author} {\bibfnamefont {R.~M.}\ \bibnamefont {Lutchyn}}, \bibinfo
  {author} {\bibfnamefont {T.~D.}\ \bibnamefont {Stanescu}},\ and\ \bibinfo
  {author} {\bibfnamefont {S.}~\bibnamefont {Das~Sarma}},\ }\bibfield  {title}
  {\bibinfo {title} {Non-{{Abelian}} quantum order in spin-orbit-coupled
  semiconductors: {{Search}} for topological {{Majorana}} particles in
  solid-state systems},\ }\href {https://doi.org/10.1103/PhysRevB.82.214509}
  {\bibfield  {journal} {\bibinfo  {journal} {Phys. Rev. B}\ }\textbf {\bibinfo
  {volume} {82}},\ \bibinfo {pages} {214509} (\bibinfo {year}
  {2010}{\natexlab{a}})}\BibitemShut {NoStop}%
\bibitem [{\citenamefont {Lutchyn}\ \emph {et~al.}(2010)\citenamefont
  {Lutchyn}, \citenamefont {Sau},\ and\ \citenamefont
  {Das~Sarma}}]{lutchyn2010majorana}%
  \BibitemOpen
  \bibfield  {author} {\bibinfo {author} {\bibfnamefont {R.~M.}\ \bibnamefont
  {Lutchyn}}, \bibinfo {author} {\bibfnamefont {J.~D.}\ \bibnamefont {Sau}},\
  and\ \bibinfo {author} {\bibfnamefont {S.}~\bibnamefont {Das~Sarma}},\
  }\bibfield  {title} {\bibinfo {title} {Majorana {{Fermions}} and a
  {{Topological Phase Transition}} in {{Semiconductor-Superconductor
  Heterostructures}}},\ }\href {https://doi.org/10.1103/PhysRevLett.105.077001}
  {\bibfield  {journal} {\bibinfo  {journal} {Phys. Rev. Lett.}\ }\textbf
  {\bibinfo {volume} {105}},\ \bibinfo {pages} {077001} (\bibinfo {year}
  {2010})}\BibitemShut {NoStop}%
\bibitem [{\citenamefont {Sau}\ \emph {et~al.}(2010{\natexlab{b}})\citenamefont
  {Sau}, \citenamefont {Lutchyn}, \citenamefont {Tewari},\ and\ \citenamefont
  {Das~Sarma}}]{sau2010generic}%
  \BibitemOpen
  \bibfield  {author} {\bibinfo {author} {\bibfnamefont {J.~D.}\ \bibnamefont
  {Sau}}, \bibinfo {author} {\bibfnamefont {R.~M.}\ \bibnamefont {Lutchyn}},
  \bibinfo {author} {\bibfnamefont {S.}~\bibnamefont {Tewari}},\ and\ \bibinfo
  {author} {\bibfnamefont {S.}~\bibnamefont {Das~Sarma}},\ }\bibfield  {title}
  {\bibinfo {title} {Generic {{New Platform}} for {{Topological Quantum
  Computation Using Semiconductor Heterostructures}}},\ }\href
  {https://doi.org/10.1103/PhysRevLett.104.040502} {\bibfield  {journal}
  {\bibinfo  {journal} {Phys. Rev. Lett.}\ }\textbf {\bibinfo {volume} {104}},\
  \bibinfo {pages} {040502} (\bibinfo {year} {2010}{\natexlab{b}})}\BibitemShut
  {NoStop}%
\bibitem [{\citenamefont {Oreg}\ \emph {et~al.}(2010)\citenamefont {Oreg},
  \citenamefont {Refael},\ and\ \citenamefont {{von Oppen}}}]{oreg2010helical}%
  \BibitemOpen
  \bibfield  {author} {\bibinfo {author} {\bibfnamefont {Y.}~\bibnamefont
  {Oreg}}, \bibinfo {author} {\bibfnamefont {G.}~\bibnamefont {Refael}},\ and\
  \bibinfo {author} {\bibfnamefont {F.}~\bibnamefont {{von Oppen}}},\
  }\bibfield  {title} {\bibinfo {title} {Helical {{Liquids}} and {{Majorana
  Bound States}} in {{Quantum Wires}}},\ }\href
  {https://doi.org/10.1103/PhysRevLett.105.177002} {\bibfield  {journal}
  {\bibinfo  {journal} {Phys. Rev. Lett.}\ }\textbf {\bibinfo {volume} {105}},\
  \bibinfo {pages} {177002} (\bibinfo {year} {2010})}\BibitemShut {NoStop}%
\bibitem [{\citenamefont {Nayak}\ \emph {et~al.}(2008)\citenamefont {Nayak},
  \citenamefont {Simon}, \citenamefont {Stern}, \citenamefont {Freedman},\ and\
  \citenamefont {Das~Sarma}}]{nayak2008nonabelian}%
  \BibitemOpen
  \bibfield  {author} {\bibinfo {author} {\bibfnamefont {C.}~\bibnamefont
  {Nayak}}, \bibinfo {author} {\bibfnamefont {S.~H.}\ \bibnamefont {Simon}},
  \bibinfo {author} {\bibfnamefont {A.}~\bibnamefont {Stern}}, \bibinfo
  {author} {\bibfnamefont {M.}~\bibnamefont {Freedman}},\ and\ \bibinfo
  {author} {\bibfnamefont {S.}~\bibnamefont {Das~Sarma}},\ }\bibfield  {title}
  {\bibinfo {title} {Non-{{Abelian}} anyons and topological quantum
  computation},\ }\href {https://doi.org/10.1103/RevModPhys.80.1083} {\bibfield
   {journal} {\bibinfo  {journal} {Reviews of Modern Physics}\ }\textbf
  {\bibinfo {volume} {80}},\ \bibinfo {pages} {1083} (\bibinfo {year}
  {2008})}\BibitemShut {NoStop}%
\bibitem [{\citenamefont {Sengupta}\ \emph {et~al.}(2001)\citenamefont
  {Sengupta}, \citenamefont {{\v Z}uti{\'c}}, \citenamefont {Kwon},
  \citenamefont {Yakovenko},\ and\ \citenamefont
  {Das~Sarma}}]{sengupta2001midgap}%
  \BibitemOpen
  \bibfield  {author} {\bibinfo {author} {\bibfnamefont {K.}~\bibnamefont
  {Sengupta}}, \bibinfo {author} {\bibfnamefont {I.}~\bibnamefont {{\v
  Z}uti{\'c}}}, \bibinfo {author} {\bibfnamefont {H.-J.}\ \bibnamefont {Kwon}},
  \bibinfo {author} {\bibfnamefont {V.~M.}\ \bibnamefont {Yakovenko}},\ and\
  \bibinfo {author} {\bibfnamefont {S.}~\bibnamefont {Das~Sarma}},\ }\bibfield
  {title} {\bibinfo {title} {Midgap edge states and pairing symmetry of
  quasi-one-dimensional organic superconductors},\ }\href
  {https://doi.org/10.1103/PhysRevB.63.144531} {\bibfield  {journal} {\bibinfo
  {journal} {Phys. Rev. B}\ }\textbf {\bibinfo {volume} {63}},\ \bibinfo
  {pages} {144531} (\bibinfo {year} {2001})}\BibitemShut {NoStop}%
\bibitem [{\citenamefont {Flensberg}(2010)}]{flensberg2010tunneling}%
  \BibitemOpen
  \bibfield  {author} {\bibinfo {author} {\bibfnamefont {K.}~\bibnamefont
  {Flensberg}},\ }\bibfield  {title} {\bibinfo {title} {Tunneling
  characteristics of a chain of {{Majorana}} bound states},\ }\href
  {https://doi.org/10.1103/PhysRevB.82.180516} {\bibfield  {journal} {\bibinfo
  {journal} {Phys. Rev. B}\ }\textbf {\bibinfo {volume} {82}},\ \bibinfo
  {pages} {180516} (\bibinfo {year} {2010})}\BibitemShut {NoStop}%
\bibitem [{\citenamefont {Wimmer}\ \emph {et~al.}(2011)\citenamefont {Wimmer},
  \citenamefont {Akhmerov}, \citenamefont {Dahlhaus},\ and\ \citenamefont
  {Beenakker}}]{wimmer2011quantum}%
  \BibitemOpen
  \bibfield  {author} {\bibinfo {author} {\bibfnamefont {M.}~\bibnamefont
  {Wimmer}}, \bibinfo {author} {\bibfnamefont {A.~R.}\ \bibnamefont
  {Akhmerov}}, \bibinfo {author} {\bibfnamefont {J.~P.}\ \bibnamefont
  {Dahlhaus}},\ and\ \bibinfo {author} {\bibfnamefont {C.~W.~J.}\ \bibnamefont
  {Beenakker}},\ }\bibfield  {title} {\bibinfo {title} {Quantum point contact
  as a probe of a topological superconductor},\ }\href
  {https://doi.org/10.1088/1367-2630/13/5/053016} {\bibfield  {journal}
  {\bibinfo  {journal} {New J. Phys.}\ }\textbf {\bibinfo {volume} {13}},\
  \bibinfo {pages} {053016} (\bibinfo {year} {2011})}\BibitemShut {NoStop}%
\bibitem [{\citenamefont {Law}\ \emph {et~al.}(2009)\citenamefont {Law},
  \citenamefont {Lee},\ and\ \citenamefont {Ng}}]{law2009majorana}%
  \BibitemOpen
  \bibfield  {author} {\bibinfo {author} {\bibfnamefont {K.~T.}\ \bibnamefont
  {Law}}, \bibinfo {author} {\bibfnamefont {P.~A.}\ \bibnamefont {Lee}},\ and\
  \bibinfo {author} {\bibfnamefont {T.~K.}\ \bibnamefont {Ng}},\ }\bibfield
  {title} {\bibinfo {title} {Majorana {{Fermion Induced Resonant Andreev
  Reflection}}},\ }\href {https://doi.org/10.1103/PhysRevLett.103.237001}
  {\bibfield  {journal} {\bibinfo  {journal} {Phys. Rev. Lett.}\ }\textbf
  {\bibinfo {volume} {103}},\ \bibinfo {pages} {237001} (\bibinfo {year}
  {2009})}\BibitemShut {NoStop}%
\bibitem [{\citenamefont {Setiawan}\ \emph {et~al.}(2017)\citenamefont
  {Setiawan}, \citenamefont {Liu}, \citenamefont {Sau},\ and\ \citenamefont
  {Das~Sarma}}]{setiawan2017electron}%
  \BibitemOpen
  \bibfield  {author} {\bibinfo {author} {\bibfnamefont {F.}~\bibnamefont
  {Setiawan}}, \bibinfo {author} {\bibfnamefont {C.-X.}\ \bibnamefont {Liu}},
  \bibinfo {author} {\bibfnamefont {J.~D.}\ \bibnamefont {Sau}},\ and\ \bibinfo
  {author} {\bibfnamefont {S.}~\bibnamefont {Das~Sarma}},\ }\bibfield  {title}
  {\bibinfo {title} {Electron temperature and tunnel coupling dependence of
  zero-bias and almost-zero-bias conductance peaks in {{Majorana}} nanowires},\
  }\href {https://doi.org/10.1103/PhysRevB.96.184520} {\bibfield  {journal}
  {\bibinfo  {journal} {Phys. Rev. B}\ }\textbf {\bibinfo {volume} {96}},\
  \bibinfo {pages} {184520} (\bibinfo {year} {2017})}\BibitemShut {NoStop}%
\bibitem [{\citenamefont {Das}\ \emph {et~al.}(2012)\citenamefont {Das},
  \citenamefont {Ronen}, \citenamefont {Most}, \citenamefont {Oreg},
  \citenamefont {Heiblum},\ and\ \citenamefont {Shtrikman}}]{das2012zerobias}%
  \BibitemOpen
  \bibfield  {author} {\bibinfo {author} {\bibfnamefont {A.}~\bibnamefont
  {Das}}, \bibinfo {author} {\bibfnamefont {Y.}~\bibnamefont {Ronen}}, \bibinfo
  {author} {\bibfnamefont {Y.}~\bibnamefont {Most}}, \bibinfo {author}
  {\bibfnamefont {Y.}~\bibnamefont {Oreg}}, \bibinfo {author} {\bibfnamefont
  {M.}~\bibnamefont {Heiblum}},\ and\ \bibinfo {author} {\bibfnamefont
  {H.}~\bibnamefont {Shtrikman}},\ }\bibfield  {title} {\bibinfo {title}
  {Zero-bias peaks and splitting in an {{Al}}\textendash{{InAs}} nanowire
  topological superconductor as a signature of {{Majorana}} fermions},\ }\href
  {https://www.nature.com/articles/nphys2479} {\bibfield  {journal} {\bibinfo
  {journal} {Nature Physics}\ }\textbf {\bibinfo {volume} {8}},\ \bibinfo
  {pages} {887} (\bibinfo {year} {2012})}\BibitemShut {NoStop}%
\bibitem [{\citenamefont {Deng}\ \emph {et~al.}(2012)\citenamefont {Deng},
  \citenamefont {Yu}, \citenamefont {Huang}, \citenamefont {Larsson},
  \citenamefont {Caroff},\ and\ \citenamefont {Xu}}]{deng2012anomalous}%
  \BibitemOpen
  \bibfield  {author} {\bibinfo {author} {\bibfnamefont {M.~T.}\ \bibnamefont
  {Deng}}, \bibinfo {author} {\bibfnamefont {C.~L.}\ \bibnamefont {Yu}},
  \bibinfo {author} {\bibfnamefont {G.~Y.}\ \bibnamefont {Huang}}, \bibinfo
  {author} {\bibfnamefont {M.}~\bibnamefont {Larsson}}, \bibinfo {author}
  {\bibfnamefont {P.}~\bibnamefont {Caroff}},\ and\ \bibinfo {author}
  {\bibfnamefont {H.~Q.}\ \bibnamefont {Xu}},\ }\bibfield  {title} {\bibinfo
  {title} {Anomalous {{Zero-Bias Conductance Peak}} in a
  {{Nb}}\textendash{{InSb Nanowire}}\textendash{{Nb Hybrid Device}}},\ }\href
  {https://doi.org/10.1021/nl303758w} {\bibfield  {journal} {\bibinfo
  {journal} {Nano Letters}\ }\textbf {\bibinfo {volume} {12}},\ \bibinfo
  {pages} {6414} (\bibinfo {year} {2012})}\BibitemShut {NoStop}%
\bibitem [{\citenamefont {Mourik}\ \emph {et~al.}(2012)\citenamefont {Mourik},
  \citenamefont {Zuo}, \citenamefont {Frolov}, \citenamefont {Plissard},
  \citenamefont {Bakkers},\ and\ \citenamefont
  {Kouwenhoven}}]{mourik2012signatures}%
  \BibitemOpen
  \bibfield  {author} {\bibinfo {author} {\bibfnamefont {V.}~\bibnamefont
  {Mourik}}, \bibinfo {author} {\bibfnamefont {K.}~\bibnamefont {Zuo}},
  \bibinfo {author} {\bibfnamefont {S.~M.}\ \bibnamefont {Frolov}}, \bibinfo
  {author} {\bibfnamefont {S.~R.}\ \bibnamefont {Plissard}}, \bibinfo {author}
  {\bibfnamefont {E.~P. A.~M.}\ \bibnamefont {Bakkers}},\ and\ \bibinfo
  {author} {\bibfnamefont {L.~P.}\ \bibnamefont {Kouwenhoven}},\ }\bibfield
  {title} {\bibinfo {title} {Signatures of {{Majorana Fermions}} in {{Hybrid
  Superconductor-Semiconductor Nanowire Devices}}},\ }\href
  {https://doi.org/10.1126/science.1222360} {\bibfield  {journal} {\bibinfo
  {journal} {Science}\ }\textbf {\bibinfo {volume} {336}},\ \bibinfo {pages}
  {1003} (\bibinfo {year} {2012})}\BibitemShut {NoStop}%
\bibitem [{\citenamefont {Churchill}\ \emph {et~al.}(2013)\citenamefont
  {Churchill}, \citenamefont {Fatemi}, \citenamefont {{Grove-Rasmussen}},
  \citenamefont {Deng}, \citenamefont {Caroff}, \citenamefont {Xu},\ and\
  \citenamefont {Marcus}}]{churchill2013superconductornanowire}%
  \BibitemOpen
  \bibfield  {author} {\bibinfo {author} {\bibfnamefont {H.~O.~H.}\
  \bibnamefont {Churchill}}, \bibinfo {author} {\bibfnamefont {V.}~\bibnamefont
  {Fatemi}}, \bibinfo {author} {\bibfnamefont {K.}~\bibnamefont
  {{Grove-Rasmussen}}}, \bibinfo {author} {\bibfnamefont {M.~T.}\ \bibnamefont
  {Deng}}, \bibinfo {author} {\bibfnamefont {P.}~\bibnamefont {Caroff}},
  \bibinfo {author} {\bibfnamefont {H.~Q.}\ \bibnamefont {Xu}},\ and\ \bibinfo
  {author} {\bibfnamefont {C.~M.}\ \bibnamefont {Marcus}},\ }\bibfield  {title}
  {\bibinfo {title} {Superconductor-nanowire devices from tunneling to the
  multichannel regime: {{Zero-bias}} oscillations and magnetoconductance
  crossover},\ }\href {https://doi.org/10.1103/PhysRevB.87.241401} {\bibfield
  {journal} {\bibinfo  {journal} {Phys. Rev. B}\ }\textbf {\bibinfo {volume}
  {87}},\ \bibinfo {pages} {241401} (\bibinfo {year} {2013})}\BibitemShut
  {NoStop}%
\bibitem [{\citenamefont {Finck}\ \emph {et~al.}(2013)\citenamefont {Finck},
  \citenamefont {Van~Harlingen}, \citenamefont {Mohseni}, \citenamefont
  {Jung},\ and\ \citenamefont {Li}}]{finck2013anomalous}%
  \BibitemOpen
  \bibfield  {author} {\bibinfo {author} {\bibfnamefont {A.~D.~K.}\
  \bibnamefont {Finck}}, \bibinfo {author} {\bibfnamefont {D.~J.}\ \bibnamefont
  {Van~Harlingen}}, \bibinfo {author} {\bibfnamefont {P.~K.}\ \bibnamefont
  {Mohseni}}, \bibinfo {author} {\bibfnamefont {K.}~\bibnamefont {Jung}},\ and\
  \bibinfo {author} {\bibfnamefont {X.}~\bibnamefont {Li}},\ }\bibfield
  {title} {\bibinfo {title} {Anomalous {{Modulation}} of a {{Zero-Bias Peak}}
  in a {{Hybrid Nanowire-Superconductor Device}}},\ }\href
  {https://doi.org/10.1103/PhysRevLett.110.126406} {\bibfield  {journal}
  {\bibinfo  {journal} {Phys. Rev. Lett.}\ }\textbf {\bibinfo {volume} {110}},\
  \bibinfo {pages} {126406} (\bibinfo {year} {2013})}\BibitemShut {NoStop}%
\bibitem [{\citenamefont {Liu}\ \emph {et~al.}(2012)\citenamefont {Liu},
  \citenamefont {Potter}, \citenamefont {Law},\ and\ \citenamefont
  {Lee}}]{liu2012zerobias}%
  \BibitemOpen
  \bibfield  {author} {\bibinfo {author} {\bibfnamefont {J.}~\bibnamefont
  {Liu}}, \bibinfo {author} {\bibfnamefont {A.~C.}\ \bibnamefont {Potter}},
  \bibinfo {author} {\bibfnamefont {K.~T.}\ \bibnamefont {Law}},\ and\ \bibinfo
  {author} {\bibfnamefont {P.~A.}\ \bibnamefont {Lee}},\ }\bibfield  {title}
  {\bibinfo {title} {Zero-{{Bias Peaks}} in the {{Tunneling Conductance}} of
  {{Spin-Orbit-Coupled Superconducting Wires}} with and without {{Majorana
  End-States}}},\ }\href {https://doi.org/10.1103/PhysRevLett.109.267002}
  {\bibfield  {journal} {\bibinfo  {journal} {Phys. Rev. Lett.}\ }\textbf
  {\bibinfo {volume} {109}},\ \bibinfo {pages} {267002} (\bibinfo {year}
  {2012})}\BibitemShut {NoStop}%
\bibitem [{\citenamefont {Bagrets}\ and\ \citenamefont
  {Altland}(2012)}]{bagrets2012class}%
  \BibitemOpen
  \bibfield  {author} {\bibinfo {author} {\bibfnamefont {D.}~\bibnamefont
  {Bagrets}}\ and\ \bibinfo {author} {\bibfnamefont {A.}~\bibnamefont
  {Altland}},\ }\bibfield  {title} {\bibinfo {title} {Class {$D$} {{Spectral
  Peak}} in {{Majorana Quantum Wires}}},\ }\href
  {https://doi.org/10.1103/PhysRevLett.109.227005} {\bibfield  {journal}
  {\bibinfo  {journal} {Phys. Rev. Lett.}\ }\textbf {\bibinfo {volume} {109}},\
  \bibinfo {pages} {227005} (\bibinfo {year} {2012})}\BibitemShut {NoStop}%
\bibitem [{\citenamefont {Akhmerov}\ \emph {et~al.}(2011)\citenamefont
  {Akhmerov}, \citenamefont {Dahlhaus}, \citenamefont {Hassler}, \citenamefont
  {Wimmer},\ and\ \citenamefont {Beenakker}}]{akhmerov2011quantized}%
  \BibitemOpen
  \bibfield  {author} {\bibinfo {author} {\bibfnamefont {A.~R.}\ \bibnamefont
  {Akhmerov}}, \bibinfo {author} {\bibfnamefont {J.~P.}\ \bibnamefont
  {Dahlhaus}}, \bibinfo {author} {\bibfnamefont {F.}~\bibnamefont {Hassler}},
  \bibinfo {author} {\bibfnamefont {M.}~\bibnamefont {Wimmer}},\ and\ \bibinfo
  {author} {\bibfnamefont {C.~W.~J.}\ \bibnamefont {Beenakker}},\ }\bibfield
  {title} {\bibinfo {title} {Quantized {{Conductance}} at the {{Majorana Phase
  Transition}} in a {{Disordered Superconducting Wire}}},\ }\href
  {https://doi.org/10.1103/PhysRevLett.106.057001} {\bibfield  {journal}
  {\bibinfo  {journal} {Phys. Rev. Lett.}\ }\textbf {\bibinfo {volume} {106}},\
  \bibinfo {pages} {057001} (\bibinfo {year} {2011})}\BibitemShut {NoStop}%
\bibitem [{\citenamefont {Sau}\ and\ \citenamefont
  {Das~Sarma}(2013)}]{sau2013density}%
  \BibitemOpen
  \bibfield  {author} {\bibinfo {author} {\bibfnamefont {J.~D.}\ \bibnamefont
  {Sau}}\ and\ \bibinfo {author} {\bibfnamefont {S.}~\bibnamefont
  {Das~Sarma}},\ }\bibfield  {title} {\bibinfo {title} {Density of states of
  disordered topological superconductor-semiconductor hybrid nanowires},\
  }\href {https://doi.org/10.1103/PhysRevB.88.064506} {\bibfield  {journal}
  {\bibinfo  {journal} {Phys. Rev. B}\ }\textbf {\bibinfo {volume} {88}},\
  \bibinfo {pages} {064506} (\bibinfo {year} {2013})}\BibitemShut {NoStop}%
\bibitem [{\citenamefont {Nichele}\ \emph {et~al.}(2017)\citenamefont
  {Nichele}, \citenamefont {Drachmann}, \citenamefont {Whiticar}, \citenamefont
  {O'Farrell}, \citenamefont {Suominen}, \citenamefont {Fornieri},
  \citenamefont {Wang}, \citenamefont {Gardner}, \citenamefont {Thomas},
  \citenamefont {Hatke}, \citenamefont {Krogstrup}, \citenamefont {Manfra},
  \citenamefont {Flensberg},\ and\ \citenamefont
  {Marcus}}]{nichele2017scaling}%
  \BibitemOpen
  \bibfield  {author} {\bibinfo {author} {\bibfnamefont {F.}~\bibnamefont
  {Nichele}}, \bibinfo {author} {\bibfnamefont {A.~C.~C.}\ \bibnamefont
  {Drachmann}}, \bibinfo {author} {\bibfnamefont {A.~M.}\ \bibnamefont
  {Whiticar}}, \bibinfo {author} {\bibfnamefont {E.~C.~T.}\ \bibnamefont
  {O'Farrell}}, \bibinfo {author} {\bibfnamefont {H.~J.}\ \bibnamefont
  {Suominen}}, \bibinfo {author} {\bibfnamefont {A.}~\bibnamefont {Fornieri}},
  \bibinfo {author} {\bibfnamefont {T.}~\bibnamefont {Wang}}, \bibinfo {author}
  {\bibfnamefont {G.~C.}\ \bibnamefont {Gardner}}, \bibinfo {author}
  {\bibfnamefont {C.}~\bibnamefont {Thomas}}, \bibinfo {author} {\bibfnamefont
  {A.~T.}\ \bibnamefont {Hatke}}, \bibinfo {author} {\bibfnamefont
  {P.}~\bibnamefont {Krogstrup}}, \bibinfo {author} {\bibfnamefont {M.~J.}\
  \bibnamefont {Manfra}}, \bibinfo {author} {\bibfnamefont {K.}~\bibnamefont
  {Flensberg}},\ and\ \bibinfo {author} {\bibfnamefont {C.~M.}\ \bibnamefont
  {Marcus}},\ }\bibfield  {title} {\bibinfo {title} {Scaling of {{Majorana
  Zero-Bias Conductance Peaks}}},\ }\href
  {https://doi.org/10.1103/PhysRevLett.119.136803} {\bibfield  {journal}
  {\bibinfo  {journal} {Phys. Rev. Lett.}\ }\textbf {\bibinfo {volume} {119}},\
  \bibinfo {pages} {136803} (\bibinfo {year} {2017})}\BibitemShut {NoStop}%
\bibitem [{\citenamefont {Zhang}\ \emph {et~al.}(2018)\citenamefont {Zhang},
  \citenamefont {Liu}, \citenamefont {Gazibegovic}, \citenamefont {Xu},
  \citenamefont {Logan}, \citenamefont {Wang}, \citenamefont {{van Loo}},
  \citenamefont {Bommer}, \citenamefont {{de Moor}}, \citenamefont {Car},
  \citenamefont {{Op het Veld}}, \citenamefont {{van Veldhoven}}, \citenamefont
  {Koelling}, \citenamefont {Verheijen}, \citenamefont {Pendharkar},
  \citenamefont {Pennachio}, \citenamefont {Shojaei}, \citenamefont {Lee},
  \citenamefont {Palmstr{\o}m}, \citenamefont {Bakkers}, \citenamefont
  {Sarma},\ and\ \citenamefont {Kouwenhoven}}]{zhang2018quantizeda}%
  \BibitemOpen
  \bibfield  {author} {\bibinfo {author} {\bibfnamefont {H.}~\bibnamefont
  {Zhang}}, \bibinfo {author} {\bibfnamefont {C.-X.}\ \bibnamefont {Liu}},
  \bibinfo {author} {\bibfnamefont {S.}~\bibnamefont {Gazibegovic}}, \bibinfo
  {author} {\bibfnamefont {D.}~\bibnamefont {Xu}}, \bibinfo {author}
  {\bibfnamefont {J.~A.}\ \bibnamefont {Logan}}, \bibinfo {author}
  {\bibfnamefont {G.}~\bibnamefont {Wang}}, \bibinfo {author} {\bibfnamefont
  {N.}~\bibnamefont {{van Loo}}}, \bibinfo {author} {\bibfnamefont {J.~D.~S.}\
  \bibnamefont {Bommer}}, \bibinfo {author} {\bibfnamefont {M.~W.~A.}\
  \bibnamefont {{de Moor}}}, \bibinfo {author} {\bibfnamefont {D.}~\bibnamefont
  {Car}}, \bibinfo {author} {\bibfnamefont {R.~L.~M.}\ \bibnamefont {{Op het
  Veld}}}, \bibinfo {author} {\bibfnamefont {P.~J.}\ \bibnamefont {{van
  Veldhoven}}}, \bibinfo {author} {\bibfnamefont {S.}~\bibnamefont {Koelling}},
  \bibinfo {author} {\bibfnamefont {M.~A.}\ \bibnamefont {Verheijen}}, \bibinfo
  {author} {\bibfnamefont {M.}~\bibnamefont {Pendharkar}}, \bibinfo {author}
  {\bibfnamefont {D.~J.}\ \bibnamefont {Pennachio}}, \bibinfo {author}
  {\bibfnamefont {B.}~\bibnamefont {Shojaei}}, \bibinfo {author} {\bibfnamefont
  {J.~S.}\ \bibnamefont {Lee}}, \bibinfo {author} {\bibfnamefont {C.~J.}\
  \bibnamefont {Palmstr{\o}m}}, \bibinfo {author} {\bibfnamefont {E.~P. A.~M.}\
  \bibnamefont {Bakkers}}, \bibinfo {author} {\bibfnamefont {S.~D.}\
  \bibnamefont {Sarma}},\ and\ \bibinfo {author} {\bibfnamefont {L.~P.}\
  \bibnamefont {Kouwenhoven}},\ }\bibfield  {title} {\bibinfo {title}
  {Quantized majorana conductance},\ }\href
  {https://doi.org/10.1038/nature26142} {\bibfield  {journal} {\bibinfo
  {journal} {[Retracted] Nature}\ }\textbf {\bibinfo {volume} {556}},\ \bibinfo
  {pages} {74} (\bibinfo {year} {2018})},\ \Eprint
  {https://arxiv.org/abs/1710.10701} {arXiv:1710.10701} \BibitemShut {NoStop}%
\bibitem [{\citenamefont {Zhang}\ \emph {et~al.}(2021)\citenamefont {Zhang},
  \citenamefont {{de Moor}}, \citenamefont {Bommer}, \citenamefont {Xu},
  \citenamefont {Wang}, \citenamefont {{van Loo}}, \citenamefont {Liu},
  \citenamefont {Gazibegovic}, \citenamefont {Logan}, \citenamefont {Car},
  \citenamefont {het Veld}, \citenamefont {{van Veldhoven}}, \citenamefont
  {Koelling}, \citenamefont {Verheijen}, \citenamefont {Pendharkar},
  \citenamefont {Pennachio}, \citenamefont {Shojaei}, \citenamefont {Lee},
  \citenamefont {Palmstr{\o}m}, \citenamefont {Bakkers}, \citenamefont
  {Sarma},\ and\ \citenamefont {Kouwenhoven}}]{zhang2021large}%
  \BibitemOpen
  \bibfield  {author} {\bibinfo {author} {\bibfnamefont {H.}~\bibnamefont
  {Zhang}}, \bibinfo {author} {\bibfnamefont {M.~W.~A.}\ \bibnamefont {{de
  Moor}}}, \bibinfo {author} {\bibfnamefont {J.~D.~S.}\ \bibnamefont {Bommer}},
  \bibinfo {author} {\bibfnamefont {D.}~\bibnamefont {Xu}}, \bibinfo {author}
  {\bibfnamefont {G.}~\bibnamefont {Wang}}, \bibinfo {author} {\bibfnamefont
  {N.}~\bibnamefont {{van Loo}}}, \bibinfo {author} {\bibfnamefont {C.-X.}\
  \bibnamefont {Liu}}, \bibinfo {author} {\bibfnamefont {S.}~\bibnamefont
  {Gazibegovic}}, \bibinfo {author} {\bibfnamefont {J.~A.}\ \bibnamefont
  {Logan}}, \bibinfo {author} {\bibfnamefont {D.}~\bibnamefont {Car}}, \bibinfo
  {author} {\bibfnamefont {R.~L. M.~O.}\ \bibnamefont {het Veld}}, \bibinfo
  {author} {\bibfnamefont {P.~J.}\ \bibnamefont {{van Veldhoven}}}, \bibinfo
  {author} {\bibfnamefont {S.}~\bibnamefont {Koelling}}, \bibinfo {author}
  {\bibfnamefont {M.~A.}\ \bibnamefont {Verheijen}}, \bibinfo {author}
  {\bibfnamefont {M.}~\bibnamefont {Pendharkar}}, \bibinfo {author}
  {\bibfnamefont {D.~J.}\ \bibnamefont {Pennachio}}, \bibinfo {author}
  {\bibfnamefont {B.}~\bibnamefont {Shojaei}}, \bibinfo {author} {\bibfnamefont
  {J.~S.}\ \bibnamefont {Lee}}, \bibinfo {author} {\bibfnamefont {C.~J.}\
  \bibnamefont {Palmstr{\o}m}}, \bibinfo {author} {\bibfnamefont {E.~P. A.~M.}\
  \bibnamefont {Bakkers}}, \bibinfo {author} {\bibfnamefont {S.~D.}\
  \bibnamefont {Sarma}},\ and\ \bibinfo {author} {\bibfnamefont {L.~P.}\
  \bibnamefont {Kouwenhoven}},\ }\bibfield  {title} {\bibinfo {title} {Large
  zero-bias peaks in {{InSb-Al}} hybrid semiconductor-superconductor nanowire
  devices},\ }\href {http://arxiv.org/abs/2101.11456} {\bibfield  {journal}
  {\bibinfo  {journal} {arXiv:2101.11456}\ } (\bibinfo {year}
  {2021})}\BibitemShut {NoStop}%
\bibitem [{\citenamefont {Yu}\ \emph {et~al.}(2021)\citenamefont {Yu},
  \citenamefont {Chen}, \citenamefont {Gomanko}, \citenamefont {Badawy},
  \citenamefont {Bakkers}, \citenamefont {Zuo}, \citenamefont {Mourik},\ and\
  \citenamefont {Frolov}}]{yu2021nonmajorana}%
  \BibitemOpen
  \bibfield  {author} {\bibinfo {author} {\bibfnamefont {P.}~\bibnamefont
  {Yu}}, \bibinfo {author} {\bibfnamefont {J.}~\bibnamefont {Chen}}, \bibinfo
  {author} {\bibfnamefont {M.}~\bibnamefont {Gomanko}}, \bibinfo {author}
  {\bibfnamefont {G.}~\bibnamefont {Badawy}}, \bibinfo {author} {\bibfnamefont
  {E.~P. a.~M.}\ \bibnamefont {Bakkers}}, \bibinfo {author} {\bibfnamefont
  {K.}~\bibnamefont {Zuo}}, \bibinfo {author} {\bibfnamefont {V.}~\bibnamefont
  {Mourik}},\ and\ \bibinfo {author} {\bibfnamefont {S.~M.}\ \bibnamefont
  {Frolov}},\ }\bibfield  {title} {\bibinfo {title} {Non-{{Majorana}} states
  yield nearly quantized conductance in proximatized nanowires},\ }\href
  {https://doi.org/10.1038/s41567-020-01107-w} {\bibfield  {journal} {\bibinfo
  {journal} {Nature Physics}\ }\textbf {\bibinfo {volume} {17}},\ \bibinfo
  {pages} {482} (\bibinfo {year} {2021})}\BibitemShut {NoStop}%
\bibitem [{\citenamefont {Pan}\ \emph {et~al.}(2020{\natexlab{a}})\citenamefont
  {Pan}, \citenamefont {Song}, \citenamefont {Zhang}, \citenamefont {Liu},
  \citenamefont {Wen}, \citenamefont {Liao}, \citenamefont {Zhuo},
  \citenamefont {Wang}, \citenamefont {Zhang}, \citenamefont {Yang},
  \citenamefont {Ying}, \citenamefont {Miao}, \citenamefont {Li}, \citenamefont
  {Shang}, \citenamefont {Zhang},\ and\ \citenamefont {Zhao}}]{pan2020situ}%
  \BibitemOpen
  \bibfield  {author} {\bibinfo {author} {\bibfnamefont {D.}~\bibnamefont
  {Pan}}, \bibinfo {author} {\bibfnamefont {H.}~\bibnamefont {Song}}, \bibinfo
  {author} {\bibfnamefont {S.}~\bibnamefont {Zhang}}, \bibinfo {author}
  {\bibfnamefont {L.}~\bibnamefont {Liu}}, \bibinfo {author} {\bibfnamefont
  {L.}~\bibnamefont {Wen}}, \bibinfo {author} {\bibfnamefont {D.}~\bibnamefont
  {Liao}}, \bibinfo {author} {\bibfnamefont {R.}~\bibnamefont {Zhuo}}, \bibinfo
  {author} {\bibfnamefont {Z.}~\bibnamefont {Wang}}, \bibinfo {author}
  {\bibfnamefont {Z.}~\bibnamefont {Zhang}}, \bibinfo {author} {\bibfnamefont
  {S.}~\bibnamefont {Yang}}, \bibinfo {author} {\bibfnamefont {J.}~\bibnamefont
  {Ying}}, \bibinfo {author} {\bibfnamefont {W.}~\bibnamefont {Miao}}, \bibinfo
  {author} {\bibfnamefont {Y.}~\bibnamefont {Li}}, \bibinfo {author}
  {\bibfnamefont {R.}~\bibnamefont {Shang}}, \bibinfo {author} {\bibfnamefont
  {H.}~\bibnamefont {Zhang}},\ and\ \bibinfo {author} {\bibfnamefont
  {J.}~\bibnamefont {Zhao}},\ }\bibfield  {title} {\bibinfo {title} {In {{Situ
  Epitaxy}} of {{Pure Phase Ultra-Thin InAs-Al Nanowires}} for {{Quantum
  Devices}}},\ }\href {http://arxiv.org/abs/2011.13620} {\bibfield  {journal}
  {\bibinfo  {journal} {arXiv:2011.13620}\ } (\bibinfo {year}
  {2020}{\natexlab{a}})}\BibitemShut {NoStop}%
\bibitem [{\citenamefont {Song}\ \emph {et~al.}(2021)\citenamefont {Song},
  \citenamefont {Zhang}, \citenamefont {Pan}, \citenamefont {Liu},
  \citenamefont {Wang}, \citenamefont {Cao}, \citenamefont {Liu}, \citenamefont
  {Wen}, \citenamefont {Liao}, \citenamefont {Zhuo}, \citenamefont {Liu},
  \citenamefont {Shang}, \citenamefont {Zhao},\ and\ \citenamefont
  {Zhang}}]{song2021large}%
  \BibitemOpen
  \bibfield  {author} {\bibinfo {author} {\bibfnamefont {H.}~\bibnamefont
  {Song}}, \bibinfo {author} {\bibfnamefont {Z.}~\bibnamefont {Zhang}},
  \bibinfo {author} {\bibfnamefont {D.}~\bibnamefont {Pan}}, \bibinfo {author}
  {\bibfnamefont {D.}~\bibnamefont {Liu}}, \bibinfo {author} {\bibfnamefont
  {Z.}~\bibnamefont {Wang}}, \bibinfo {author} {\bibfnamefont {Z.}~\bibnamefont
  {Cao}}, \bibinfo {author} {\bibfnamefont {L.}~\bibnamefont {Liu}}, \bibinfo
  {author} {\bibfnamefont {L.}~\bibnamefont {Wen}}, \bibinfo {author}
  {\bibfnamefont {D.}~\bibnamefont {Liao}}, \bibinfo {author} {\bibfnamefont
  {R.}~\bibnamefont {Zhuo}}, \bibinfo {author} {\bibfnamefont {D.~E.}\
  \bibnamefont {Liu}}, \bibinfo {author} {\bibfnamefont {R.}~\bibnamefont
  {Shang}}, \bibinfo {author} {\bibfnamefont {J.}~\bibnamefont {Zhao}},\ and\
  \bibinfo {author} {\bibfnamefont {H.}~\bibnamefont {Zhang}},\ }\bibfield
  {title} {\bibinfo {title} {Large zero bias peaks and dips in a four-terminal
  thin {{InAs-Al}} nanowire device},\ }\href {http://arxiv.org/abs/2107.08282}
  {\bibfield  {journal} {\bibinfo  {journal} {arXiv:2107.08282}\ } (\bibinfo
  {year} {2021})}\BibitemShut {NoStop}%
\bibitem [{\citenamefont {Das~Sarma}\ \emph {et~al.}(2012)\citenamefont
  {Das~Sarma}, \citenamefont {Sau},\ and\ \citenamefont
  {Stanescu}}]{dassarma2012splitting}%
  \BibitemOpen
  \bibfield  {author} {\bibinfo {author} {\bibfnamefont {S.}~\bibnamefont
  {Das~Sarma}}, \bibinfo {author} {\bibfnamefont {J.~D.}\ \bibnamefont {Sau}},\
  and\ \bibinfo {author} {\bibfnamefont {T.~D.}\ \bibnamefont {Stanescu}},\
  }\bibfield  {title} {\bibinfo {title} {Splitting of the zero-bias conductance
  peak as smoking gun evidence for the existence of the {{Majorana}} mode in a
  superconductor-semiconductor nanowire},\ }\href
  {https://doi.org/10.1103/PhysRevB.86.220506} {\bibfield  {journal} {\bibinfo
  {journal} {Phys. Rev. B}\ }\textbf {\bibinfo {volume} {86}},\ \bibinfo
  {pages} {220506} (\bibinfo {year} {2012})}\BibitemShut {NoStop}%
\bibitem [{\citenamefont {Pan}\ and\ \citenamefont
  {Das~Sarma}(2020)}]{pan2020physical}%
  \BibitemOpen
  \bibfield  {author} {\bibinfo {author} {\bibfnamefont {H.}~\bibnamefont
  {Pan}}\ and\ \bibinfo {author} {\bibfnamefont {S.}~\bibnamefont
  {Das~Sarma}},\ }\bibfield  {title} {\bibinfo {title} {Physical mechanisms for
  zero-bias conductance peaks in {{Majorana}} nanowires},\ }\href
  {https://doi.org/10.1103/PhysRevResearch.2.013377} {\bibfield  {journal}
  {\bibinfo  {journal} {Phys. Rev. Research}\ }\textbf {\bibinfo {volume}
  {2}},\ \bibinfo {pages} {013377} (\bibinfo {year} {2020})}\BibitemShut
  {NoStop}%
\bibitem [{\citenamefont {Pan}\ \emph {et~al.}(2020{\natexlab{b}})\citenamefont
  {Pan}, \citenamefont {Cole}, \citenamefont {Sau},\ and\ \citenamefont
  {Das~Sarma}}]{pan2020generic}%
  \BibitemOpen
  \bibfield  {author} {\bibinfo {author} {\bibfnamefont {H.}~\bibnamefont
  {Pan}}, \bibinfo {author} {\bibfnamefont {W.~S.}\ \bibnamefont {Cole}},
  \bibinfo {author} {\bibfnamefont {J.~D.}\ \bibnamefont {Sau}},\ and\ \bibinfo
  {author} {\bibfnamefont {S.}~\bibnamefont {Das~Sarma}},\ }\bibfield  {title}
  {\bibinfo {title} {Generic quantized zero-bias conductance peaks in
  superconductor-semiconductor hybrid structures},\ }\href
  {https://doi.org/10.1103/PhysRevB.101.024506} {\bibfield  {journal} {\bibinfo
   {journal} {Phys. Rev. B}\ }\textbf {\bibinfo {volume} {101}},\ \bibinfo
  {pages} {024506} (\bibinfo {year} {2020}{\natexlab{b}})}\BibitemShut
  {NoStop}%
\bibitem [{\citenamefont {Pan}\ and\ \citenamefont
  {Das~Sarma}(2021{\natexlab{a}})}]{pan2021disorder}%
  \BibitemOpen
  \bibfield  {author} {\bibinfo {author} {\bibfnamefont {H.}~\bibnamefont
  {Pan}}\ and\ \bibinfo {author} {\bibfnamefont {S.}~\bibnamefont
  {Das~Sarma}},\ }\bibfield  {title} {\bibinfo {title} {Disorder effects on
  {{Majorana}} zero modes: {{Kitaev}} chain versus semiconductor nanowire},\
  }\href {https://doi.org/10.1103/PhysRevB.103.224505} {\bibfield  {journal}
  {\bibinfo  {journal} {Phys. Rev. B}\ }\textbf {\bibinfo {volume} {103}},\
  \bibinfo {pages} {224505} (\bibinfo {year} {2021}{\natexlab{a}})}\BibitemShut
  {NoStop}%
\bibitem [{\citenamefont {Pan}\ \emph {et~al.}(2021{\natexlab{a}})\citenamefont
  {Pan}, \citenamefont {Sau},\ and\ \citenamefont
  {Das~Sarma}}]{pan2021threeterminal}%
  \BibitemOpen
  \bibfield  {author} {\bibinfo {author} {\bibfnamefont {H.}~\bibnamefont
  {Pan}}, \bibinfo {author} {\bibfnamefont {J.~D.}\ \bibnamefont {Sau}},\ and\
  \bibinfo {author} {\bibfnamefont {S.}~\bibnamefont {Das~Sarma}},\ }\bibfield
  {title} {\bibinfo {title} {Three-terminal nonlocal conductance in
  {{Majorana}} nanowires: {{Distinguishing}} topological and trivial in
  realistic systems with disorder and inhomogeneous potential},\ }\href
  {https://doi.org/10.1103/PhysRevB.103.014513} {\bibfield  {journal} {\bibinfo
   {journal} {Phys. Rev. B}\ }\textbf {\bibinfo {volume} {103}},\ \bibinfo
  {pages} {014513} (\bibinfo {year} {2021}{\natexlab{a}})}\BibitemShut
  {NoStop}%
\bibitem [{\citenamefont {Pan}\ and\ \citenamefont
  {Das~Sarma}(2021{\natexlab{b}})}]{pan2021crossover}%
  \BibitemOpen
  \bibfield  {author} {\bibinfo {author} {\bibfnamefont {H.}~\bibnamefont
  {Pan}}\ and\ \bibinfo {author} {\bibfnamefont {S.}~\bibnamefont
  {Das~Sarma}},\ }\bibfield  {title} {\bibinfo {title} {Crossover between
  trivial zero modes in {{Majorana}} nanowires},\ }\href
  {https://doi.org/10.1103/PhysRevB.104.054510} {\bibfield  {journal} {\bibinfo
   {journal} {Phys. Rev. B}\ }\textbf {\bibinfo {volume} {104}},\ \bibinfo
  {pages} {054510} (\bibinfo {year} {2021}{\natexlab{b}})}\BibitemShut
  {NoStop}%
\bibitem [{\citenamefont {Pan}\ \emph {et~al.}(2021{\natexlab{b}})\citenamefont
  {Pan}, \citenamefont {Liu}, \citenamefont {Wimmer},\ and\ \citenamefont
  {Das~Sarma}}]{pan2021quantized}%
  \BibitemOpen
  \bibfield  {author} {\bibinfo {author} {\bibfnamefont {H.}~\bibnamefont
  {Pan}}, \bibinfo {author} {\bibfnamefont {C.-X.}\ \bibnamefont {Liu}},
  \bibinfo {author} {\bibfnamefont {M.}~\bibnamefont {Wimmer}},\ and\ \bibinfo
  {author} {\bibfnamefont {S.}~\bibnamefont {Das~Sarma}},\ }\bibfield  {title}
  {\bibinfo {title} {Quantized and unquantized zero-bias tunneling conductance
  peaks in {{Majorana}} nanowires: {{Conductance}} below and above
  {$2{e}^{2}/h$}},\ }\href {https://doi.org/10.1103/PhysRevB.103.214502}
  {\bibfield  {journal} {\bibinfo  {journal} {Phys. Rev. B}\ }\textbf {\bibinfo
  {volume} {103}},\ \bibinfo {pages} {214502} (\bibinfo {year}
  {2021}{\natexlab{b}})}\BibitemShut {NoStop}%
\bibitem [{\citenamefont {Das~Sarma}\ and\ \citenamefont
  {Pan}(2021)}]{dassarma2021disorderinduced}%
  \BibitemOpen
  \bibfield  {author} {\bibinfo {author} {\bibfnamefont {S.}~\bibnamefont
  {Das~Sarma}}\ and\ \bibinfo {author} {\bibfnamefont {H.}~\bibnamefont
  {Pan}},\ }\bibfield  {title} {\bibinfo {title} {Disorder-induced zero-bias
  peaks in {{Majorana}} nanowires},\ }\href
  {https://doi.org/10.1103/PhysRevB.103.195158} {\bibfield  {journal} {\bibinfo
   {journal} {Phys. Rev. B}\ }\textbf {\bibinfo {volume} {103}},\ \bibinfo
  {pages} {195158} (\bibinfo {year} {2021})}\BibitemShut {NoStop}%
\bibitem [{\citenamefont {Woods}\ \emph {et~al.}(2021)\citenamefont {Woods},
  \citenamefont {Sarma},\ and\ \citenamefont {Stanescu}}]{woods2021charge}%
  \BibitemOpen
  \bibfield  {author} {\bibinfo {author} {\bibfnamefont {B.~D.}\ \bibnamefont
  {Woods}}, \bibinfo {author} {\bibfnamefont {S.~D.}\ \bibnamefont {Sarma}},\
  and\ \bibinfo {author} {\bibfnamefont {T.~D.}\ \bibnamefont {Stanescu}},\
  }\bibfield  {title} {\bibinfo {title} {Charge impurity effects in hybrid
  {{Majorana}} nanowires},\ }\href {http://arxiv.org/abs/2103.06880} {\bibfield
   {journal} {\bibinfo  {journal} {arXiv:2103.06880}\ } (\bibinfo {year}
  {2021})}\BibitemShut {NoStop}%
\bibitem [{\citenamefont {Zeng}\ \emph {et~al.}(2021)\citenamefont {Zeng},
  \citenamefont {Sharma}, \citenamefont {Tewari},\ and\ \citenamefont
  {Stanescu}}]{zeng2021partiallyseparated}%
  \BibitemOpen
  \bibfield  {author} {\bibinfo {author} {\bibfnamefont {C.}~\bibnamefont
  {Zeng}}, \bibinfo {author} {\bibfnamefont {G.}~\bibnamefont {Sharma}},
  \bibinfo {author} {\bibfnamefont {S.}~\bibnamefont {Tewari}},\ and\ \bibinfo
  {author} {\bibfnamefont {T.}~\bibnamefont {Stanescu}},\ }\bibfield  {title}
  {\bibinfo {title} {Partially-separated {{Majorana}} modes in a disordered
  medium},\ }\href {http://arxiv.org/abs/2105.06469} {\bibfield  {journal}
  {\bibinfo  {journal} {arXiv:2105.06469}\ } (\bibinfo {year}
  {2021})}\BibitemShut {NoStop}%
\bibitem [{\citenamefont {Ahn}\ \emph {et~al.}(2021)\citenamefont {Ahn},
  \citenamefont {Pan}, \citenamefont {Woods}, \citenamefont {Stanescu},\ and\
  \citenamefont {Sarma}}]{ahn2021estimating}%
  \BibitemOpen
  \bibfield  {author} {\bibinfo {author} {\bibfnamefont {S.}~\bibnamefont
  {Ahn}}, \bibinfo {author} {\bibfnamefont {H.}~\bibnamefont {Pan}}, \bibinfo
  {author} {\bibfnamefont {B.}~\bibnamefont {Woods}}, \bibinfo {author}
  {\bibfnamefont {T.~D.}\ \bibnamefont {Stanescu}},\ and\ \bibinfo {author}
  {\bibfnamefont {S.~D.}\ \bibnamefont {Sarma}},\ }\bibfield  {title} {\bibinfo
  {title} {Estimating disorder and its adverse effects in semiconductor
  {{Majorana}} nanowires},\ }\href {http://arxiv.org/abs/2109.00007} {\bibfield
   {journal} {\bibinfo  {journal} {arXiv:2109.00007}\ } (\bibinfo {year}
  {2021})}\BibitemShut {NoStop}%
\bibitem [{\citenamefont {Groth}\ \emph {et~al.}(2014)\citenamefont {Groth},
  \citenamefont {Wimmer}, \citenamefont {Akhmerov},\ and\ \citenamefont
  {Waintal}}]{groth2014kwant}%
  \BibitemOpen
  \bibfield  {author} {\bibinfo {author} {\bibfnamefont {C.~W.}\ \bibnamefont
  {Groth}}, \bibinfo {author} {\bibfnamefont {M.}~\bibnamefont {Wimmer}},
  \bibinfo {author} {\bibfnamefont {A.~R.}\ \bibnamefont {Akhmerov}},\ and\
  \bibinfo {author} {\bibfnamefont {X.}~\bibnamefont {Waintal}},\ }\bibfield
  {title} {\bibinfo {title} {Kwant: A software package for quantum transport},\
  }\href {http://iopscience.iop.org/article/10.1088/1367-2630/16/6/063065/meta}
  {\bibfield  {journal} {\bibinfo  {journal} {New Journal of Physics}\ }\textbf
  {\bibinfo {volume} {16}},\ \bibinfo {pages} {063065} (\bibinfo {year}
  {2014})}\BibitemShut {NoStop}%
\bibitem [{\citenamefont {Huang}\ \emph {et~al.}(2018)\citenamefont {Huang},
  \citenamefont {Pan}, \citenamefont {Liu}, \citenamefont {Sau}, \citenamefont
  {Stanescu},\ and\ \citenamefont {Das~Sarma}}]{huang2018metamorphosis}%
  \BibitemOpen
  \bibfield  {author} {\bibinfo {author} {\bibfnamefont {Y.}~\bibnamefont
  {Huang}}, \bibinfo {author} {\bibfnamefont {H.}~\bibnamefont {Pan}}, \bibinfo
  {author} {\bibfnamefont {C.-X.}\ \bibnamefont {Liu}}, \bibinfo {author}
  {\bibfnamefont {J.~D.}\ \bibnamefont {Sau}}, \bibinfo {author} {\bibfnamefont
  {T.~D.}\ \bibnamefont {Stanescu}},\ and\ \bibinfo {author} {\bibfnamefont
  {S.}~\bibnamefont {Das~Sarma}},\ }\bibfield  {title} {\bibinfo {title}
  {Metamorphosis of {{Andreev}} bound states into {{Majorana}} bound states in
  pristine nanowires},\ }\href {https://doi.org/10.1103/PhysRevB.98.144511}
  {\bibfield  {journal} {\bibinfo  {journal} {Phys. Rev. B}\ }\textbf {\bibinfo
  {volume} {98}},\ \bibinfo {pages} {144511} (\bibinfo {year}
  {2018})}\BibitemShut {NoStop}%
\bibitem [{\citenamefont {Blonder}\ \emph {et~al.}(1982)\citenamefont
  {Blonder}, \citenamefont {Tinkham},\ and\ \citenamefont
  {Klapwijk}}]{blonder1982transition}%
  \BibitemOpen
  \bibfield  {author} {\bibinfo {author} {\bibfnamefont {G.~E.}\ \bibnamefont
  {Blonder}}, \bibinfo {author} {\bibfnamefont {M.}~\bibnamefont {Tinkham}},\
  and\ \bibinfo {author} {\bibfnamefont {T.~M.}\ \bibnamefont {Klapwijk}},\
  }\bibfield  {title} {\bibinfo {title} {Transition from metallic to tunneling
  regimes in superconducting microconstrictions: {{Excess}} current, charge
  imbalance, and supercurrent conversion},\ }\href
  {https://doi.org/10.1103/PhysRevB.25.4515} {\bibfield  {journal} {\bibinfo
  {journal} {Phys. Rev. B}\ }\textbf {\bibinfo {volume} {25}},\ \bibinfo
  {pages} {4515} (\bibinfo {year} {1982})}\BibitemShut {NoStop}%
\end{thebibliography}%

\appendix
\section{Theory of the superconductor-semiconductor nanowire}\label{App:A}
In this section, we briefly introduce the numerical details of the superconductor-semiconductor nanowire used in the main text to calculate the tunnel conductance. The Hamiltonian for a single subband one-dimensional model with a length of $L$ in the Bogoliubov de-Gennes Hamiltonian is $\hat{H}=\frac{1}{2}\int_0^L dx \hat{\Psi}^\dagger(x) H_{\text{BdG}}(x) \hat{\Psi}(x)$, where {the Nambu basis} $\hat{\Psi}(x)=\left(\hat{\psi}_\uparrow(x),\hat{\psi}_\downarrow(x),\hat{\psi}_\downarrow(x)^\dagger, -\hat{\psi}_\uparrow(x)^\dagger\right)^\intercal$, and
\begin{widetext}
\begin{equation}\label{eq:Ham}
    H_{\text{BdG}}(x)=\left( -\frac{\hbar^2\partial_x^2}{2m^*} -i \alpha \partial_x \sigma_y -\mu \right) \tau_z +V_z \sigma_x -\gamma \frac{\omega+\Delta_0 \tau_x}{\sqrt{\Delta_0^2-\omega^2}}+V_{\text{dis}}(x)\tau_z.
\end{equation}
\end{widetext}
Here, the first term describes the pristine semiconductor, where $\alpha$ is the Rashba-type spin-orbit coupling, $m^*$ is the effective mass for the conduction band, $\mu$ is the constant chemical potential relative to the band bottom, and $\vec{\sigma}$ ($\vec{\tau}$) are the Pauli matrices acting on the spin (particle-hole) space. The second term describes the Zeeman field. The third term represents the self-energy of the superconductor which induces the proximitized superconductivity to the semiconductor, where $\Delta_0$ is the parent superconductivity and $\gamma$ is the superconductor-semiconductor coupling strength. The last term describes the potential induced by random disorder, where we manually set the disorder magnitude very large in the bulk of nanowire while weak near two edges due to the screening by the metallic gates ({$V_{\text{dis}}$ is zero for the pristine wire}).  

To calculate the tunnel conductance of the nanowire, we follow the Blonder-Tinkham-Kalpwijk formalism~\cite{blonder1982transition} and attach the normal lead to the end of the nanowire~\cite{setiawan2017electron}. The Hamiltonian for the normal lead is the same as the Hamiltonian for the pristine semiconductor in the presence of the Zeeman field [i.e., the first plus the second term in Eq.~\eqref{eq:Ham}] with the only difference being in the chemical potential $\mu$ since the Fermi energy in the normal metal is much higher than that in the semiconductor, where we choose $\mu_{\text{Lead}}=25$ meV as opposed to $\mu_{\text{SM}}=1$ meV in the nanowire. At the interface between the normal metal and the nanowire, we model the tunnel barrier in the form of a Dirac $\delta$ function, which corresponds to the tunnel gate voltage in experiments. Namely, an additional term $V_{b}\delta(x)$ is added to Eq.~\eqref{eq:Ham} at the normal metal-superconductor junction. Here, the barrier height $V_b$ inversely characterizes the tunnel coupling strength: a high (low) $V_b$ corresponds to a low (high) coupling strength and so does for the tunnel transparency between the normal lead and the nanowire. In our simulation, we vary $V_b$ from 5 to 20 meV to change from a high-transparency limit to a low-transparency limit.

\section{Local density of states and wave functions in nanowire}\label{App:B}

\begin{figure*}[ht]
    \centering
    \includegraphics[width=6.8in]{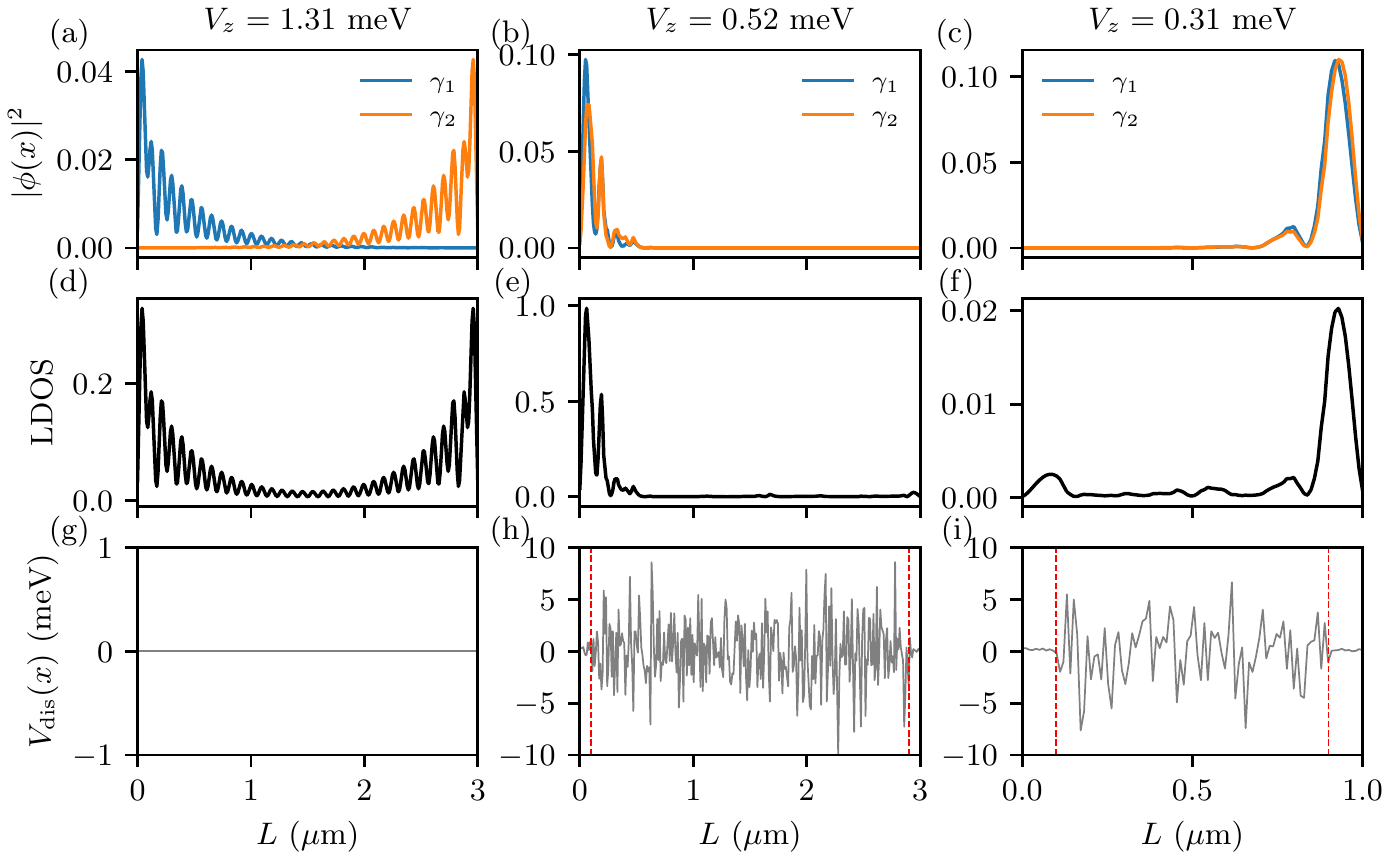}
    \caption{(a)-(c) The wave functions in the Majorana basis at a specific $V_z$ for the pristine wire, the long disordered wire, and the short disordered wire following Figs.~\ref{fig:1}-\ref{fig:4}, respectively. (d)-(f) The corresponding LDOS at zero energy. (g)-(i) The corresponding disorder spatial profiles. The red dashed lines indicate the wire edges where the disorder is suppressed due to the screening of metallic leads.}
    \label{fig:App1}
\end{figure*}

In this section, we calculate the local density of states (LDOS) and wave functions in the nanowire for three aforementioned cases (the pristine wire in the first column, the long disordered wire in the second column, and the short disordered wire in the third column) to explicitly show that the Andreev bound states are localized in the effective quantum dots at the wire ends due to the screening of metallic leads, which leads to the possibility of the large-conductance of zero-bias peaks under small tunnel barriers (large tunnel amplitude).

The wave functions shown in the first row of Fig.~\ref{fig:App1} are obtained by decomposing the wave functions in the Nambu basis $\hat{\Psi}(x)$ to the Majorana basis, namely,
\begin{eqnarray}
    \phi_{1,E_n}(x)&=&\frac{1}{\sqrt{2}}\left( \hat{\Psi}_{E_n}(x)+\hat{\Psi}_{-E_n}(x) \right),\\
    \phi_{2,E_n}(x)&=&\frac{i}{\sqrt{2}}\left( \hat{\Psi}_{E_n}(x)-\hat{\Psi}_{-E_n}(x) \right).
\end{eqnarray}

Here, $\phi_{1,2}$ are two wave functions in the basis of two Majorana operators $\gamma_{1,2}$ ($c^\dagger=\gamma_1-i\gamma_2$), and $E_n$ denotes the energy of the $n$th state, where we choose the lowest state to plot the wave functions.

The LDOS shown in the second row of Fig.~\ref{fig:App1} is defined as
\begin{equation}
    \text{LDOS}(\omega,x_i)=-\frac{1}{\pi}  \Im\left[\tr_{\sigma,\tau}G (\omega,H_{\text{BdG}}) \right]_{i,i},
\end{equation}
where  $G (\omega,H_{\text{BdG}})=\left( \omega+\delta_0-H_{\text{BdG}} \right)^{-1}$ is the Green's function of $H_{\text{BdG}}$ at the energy $\omega$, and $\delta_0$ is small for the inverse of lifetime; $\Im(\dots)$ takes the imaginary part; and $\tr_{\sigma,\tau}(\dots)$ denotes the partial trace over the spin space ($\sigma$) and the particle-hole space ($\tau$). We take the $i$th entry of the diagonal term to obtain the LDOS at the position $x_i$. In Figs.~\ref{fig:App1}(d)-(f), we show the LDOS along the wire at a specific Zeeman field and zero energy ($\omega=0$) .

In Fig.~\ref{fig:App1}, the first column corresponds to the pristine nanowire (Fig.~\ref{fig:1}), the second column corresponds to the long disordered nanowire (Figs.~\ref{fig:2} and \ref{fig:3}), and the third column corresponds to the short disordered nanowire (Fig.~\ref{fig:4}). In the pristine case without any disorder, the ZBCPs are created by a pair of topological MZMs localized at both ends. However, in the long disordered wire, the two ends of the wire manifest much smaller disorder than the bulk region, which serves as effective quantum dots [denoted by the red dashed lines in Figs.~\ref{fig:App1}(h) and \ref{fig:App1}(i)]. The two Majorana modes are both localized at the left end of the wire, which indicates trivial Andreev bound states that manifest the large conductance zero-bias peaks in Fig.~\ref{fig:2}. Similarly, the short wire also traps a pair of Majorana at the right end, which leads to a large conductance zero-bias peak.

\end{document}